\documentclass[12pt]{article}
\usepackage{amssymb}
\usepackage{cite}
\usepackage[]{hyperref}
\input xy
\xyoption{all}

\begin{document}

\hfill UWThPh2017-38

\vspace*{0.5in}

\begin{center}

{\large\bf GLSM realizations of maps \\ 
and intersections of Grassmannians and Pfaffians}

\vspace{0.25in}

Andrei C\u{a}ld\u{a}raru$^1$, Johanna Knapp$^2$, Eric Sharpe$^3$ \\

\vspace*{0.2in}

\begin{tabular}{ccc}
{ \begin{tabular}{l}
$^1$ Dep't of Mathematics\\
University of Wisconsin\\
480 Lincoln Dr.\\
Madison, WI  53706
\end{tabular} } &
{ \begin{tabular}{l}
$^2$ Mathematical Physics Group\\
University of Vienna\\
Boltzmanngasse 5\\
1090 Vienna, 
Austria
\end{tabular} } &
{ \begin{tabular}{l}
$^3$ Dep't of Physics\\
Virginia Tech\\
850 West Campus Dr.\\
Blacksburg, VA  24061
\end{tabular} }
\end{tabular}

{\tt andreic@math.wisc.edu},
{\tt johanna.knapp@univie.ac.at},
{\tt ersharpe@vt.edu}

$\,$

\end{center}

In this paper we give gauged linear sigma model (GLSM) 
realizations of a number of geometries not
previously presented in GLSMs.
We begin by describing GLSM realizations of maps including
Veronese and Segre embeddings, which can be applied to give GLSMs
explicitly describing constructions such as the intersection of one
hypersurface with the image under some map of another.
We also 
discuss GLSMs for intersections of Grassmannians and Pfaffians with
one another, and with their images under various maps, which sometimes
form exotic constructions of Calabi-Yaus, as well as GLSMs for
other exotic Calabi-Yau constructions of Kanazawa.
Much of this paper focuses on a specific set of examples of
GLSMs for intersections of Grassmannians $G(2,N)$ with themselves after a linear
rotation,
including the Calabi-Yau case $N=5$. 
One phase of the GLSM realizes an intersection of
two Grassmannians, the other phase realizes an intersection of two
Pfaffians.  The GLSM has two nonabelian factors in its gauge group,
and we consider dualities in those factors.
In both the original GLSM and a double-dual, one geometric
phase is realized perturbatively (as the critical locus of a superpotential),
and the other via quantum effects.  Dualizing on a single gauge group factor
yields a model in which each geometry is realized through a simultaneous
combination of perturbative and quantum effects.

\begin{flushleft}
October 2017
\end{flushleft}

\newpage

\tableofcontents

\newpage

\section{Introduction}

Over the last decade, a number of works have appeared showing how to construct
GLSMs for more exotic spaces than complete intersections in toric
varieties, in both abelian and nonabelian theories.  For example, 
there now exist constructions for Pfaffian varieties, both
as perturbative critical loci as well as with nonperturbative methods, 
nonperturbative
constructions of branched double covers and noncommutative resolutions
thereof, and GLSMs in which the geometric phases are not birational.
Even though one expects a whole zoo of new examples, 
only a handful of explicit constructions have been given so far 
\cite{Hori:2006dk,Donagi:2007hi,Caldararu:2007tc,Hori:2011pd,Jockers:2012zr,Hori:2013gga,Gerhardus:2015sla,Gerhardus:2016iot,Hori:2016txh}.

It is obvious that a greater variety of examples should exist. For instance, besides the Pl\"ucker embedding that plays a central role in the Grassmannian examples, there are further embeddings such as the Veronese embedding and the Segre embedding. Part of the motivation of this work was to give a realization of these embeddings in terms of the GLSM. 

A further motivation comes from recent constructions of exotic Calabi-Yau geometries in the mathematics literaure that naturally generalize the known Grassmannian and Pfaffian constructions. One of these is a complete intersection of two Grassmannians in $\mathbb{P}^9$ and its homological projective dual \cite{kan} that has recently been shown to be a counter example to the birational Torelli problem \cite{bcp,or}. In this work we give a GLSM realization of this pair of non-birational one-parameter Calabi-Yaus. We furtermore give a GLSM description of two Calabi-Yaus that are intersections of Pfaffian varieties with hypersurfaces \cite{kan}.

We begin in section~\ref{sect:maps} by describing GLSMs for (graphs of)
functions such as Veronese and Segre embeddings.  For example, the image
of the Veronese embedding of degree two is a determinantal variety, and we
will see that the GLSM for the Veronese duplicates the PAXY model
\cite{Jockers:2012zr} for that determinantal variety.
In section~\ref{sect:int-nonabelian} we discuss GLSMs for various
intersections of
Grassmannians and Pfaffians.
In section~\ref{sect:phases} we study phases of the GLSM for an 
intersection of Grassmannians, giving a physical realization of
\cite{bcp,or}.  The $r \ll 0$ phase of this GLSM realizes
an intersection of Pfaffians, in a nonperturbative fashion, and we apply
dualities which
yield GLSMs in which the phases are realized in a variety of other fashions.
Finally in section~\ref{sect:kanazawa-exs} we examine GLSMs for other exotic
mathematical constructions of Calabi-Yaus discussed in 
\cite{kan}.

\section{GLSM realizations of various maps}
\label{sect:maps}

\subsection{Veronese embeddings}

Recall a Veronese map of degree $d$
is a map $\nu: {\mathbb P} V \rightarrow{\mathbb P}( {\rm Sym}^d V)$
that sends homogeneous coordinates $[x_0, \cdots, x_n]$ (on an $n+1$-dimensional
vector space $V$) to all possible
monomials of degree $d$.
For example, the degree two map ${\mathbb P}^1 \rightarrow {\mathbb P}^2$
sends
\begin{displaymath}
[x_0, x_1] \: \mapsto \: [x_0^2, x_1^2, x_0 x_1].
\end{displaymath}
If ${\mathbb P}(V)$ is a projective space of dimension $n$,
then ${\mathbb P}({\rm Sym}^d V)$ is a projective space of dimension
\begin{displaymath}
\left( \begin{array}{c} n+d \\ d \end{array} \right) -1.
\end{displaymath}

We can realize Veronese maps in GLSMs 
by adding a series of Lagrange multipliers to identify
homogeneous coordinates in the image projective space with monomials
of homogeneous coordinates in the original projective space.  
By integrating out the `target' homogeneous coordinates, we are left
with the original projective space.  Conversely, if we can integrate out
the homogeneous coordinates of the original projective space, we are
left with a realization of the image on the target projective space.
Such images form examples of (hyper)determinantal varieties, and in fact
we shall see later that this construction is equivalent to the PAXY model of
\cite{Jockers:2012zr} for such (hyper)determinantal varieties.
In this fashion, we realize the Veronese embedding in a GLSM, or perhaps
more accurately, its graph.

For example, consider the degree
two map ${\mathbb P}^1 \rightarrow {\mathbb P}^2$ above.
Corresponding to this map, consider a GLSM defined by
a $U(1)$ gauge theory with chiral superfields\footnote{By a slight abuse of notation we use the same letters to denote superfields and their scalar components. }
$x_{0, 1}$, of charge $1$, $y_{0, 1, 2}$ of charge $2$,
and $p_{0, 1, 2}$ of charge $-2$, with superpotential
\begin{displaymath}
W \: = \: p_0 ( y_0 - x_0^2 ) \: + \:
p_1 ( y_1 - x_1^2 ) \: + \:
p_2 ( y_2 - x_0 x_1 ).
\end{displaymath}
The D-terms require that the collection $\{ x_0, \cdots, y_2 \}$
not be all zero in the phase in which $r\gg0$, 
where $r$ is the FI parameter.  The F-terms require
\begin{displaymath}
y_i = \nu_i(x), \: \: \:
p_i = 0, \: \: \: 
\sum_i p_i \partial_j \nu_i(x) = 0,
\end{displaymath}
where $\nu_i(x)$ denotes the image of the Veronese map.
Via the F terms,
the $p$'s act as Lagrange multipliers, forcing $y_i$ to be the image
of the Veronese embedding.
Furthermore, the F terms for the $y_i$'s immediately require
all $p_i=0$, as expected to realize the desired geometry.
(Mathematically, the Veronese embedding is always smooth,
see {\it e.g.} \cite{gh}[chapter 1.4],
which is also consistent
with the observation that the $p_i$'s never become nonzero on vacua.)

Integrating out $y$'s and $p$'s reduces this model to just ${\mathbb P}^1$,
realized with $x$'s.  (We will see that this is a special case of 
a more general result, for GLSM's realizing functions, or more properly
graphs of functions, in section~\ref{sect:general-maps}.)

More properly, this GLSM is realizing a locus in
${\mathbb P}^4_{[1,1,2,2,2]}$, where the last three homogeneous coordinates
are identified with quadratic monomials in the first two.  However, this
locus is the same as a ${\mathbb P}^1$, parametrized by the first two
homogeneous coordinates.  As a quick consistency check, note that those
first two homogeneous coordinates cannot both vanish:  if they did,
the other homogeneous coordinates would immediately vanish, and that point
does not exist in the weighted projective space.  For the same reason,
this locus does not intersect the ${\mathbb Z}_2$ singularities of the
weighted projective space, which arise where the first two homogeneous 
coordinates vanish.

Also note that if we define
\begin{displaymath}
p_{00} = p_0, \: \: \:
p_{11} = p_1, \: \: \:
p_{10} = p_{01} = p_2,
\end{displaymath}
\begin{displaymath}
y_{00} = y_0, \: \: \:
y_{11} = y_1, \: \: \:
y_{10} = y_{01} = y_2,
\end{displaymath}
then we can rewrite the potential in the form
\begin{displaymath}
W \: = \: \sum_{i < j} p_{ij}\left( y_{ij} - x_i x_j \right),
\end{displaymath}
which has the same form as 
a symmetric-matrix version of the PAXY model discussed
in \cite{Jockers:2012zr}[section 3.5].

Let us take a moment to 
compare in greater detail to the pertinent PAXY model, for symmetric
matrices, discussed in \cite{Jockers:2012zr}[section 3.5].
There, to describe the locus on which a symmetric $n \times n$
matrix $A(\phi)_{ij}$ has rank $k$, they introduced chiral superfields
$p_{ij}$, an $O(k)$ gauge symmetry, a set of $n$ chiral superfields $x_{ia}$
valued in the fundamental, and a superpotential
\begin{displaymath}
W \: = \: p_{ij} \left( A(\phi)_{ij} - x_{i a} x_{j a} \right).
\end{displaymath}

In the present case, to describe the image of the Veronese embedding,
we want to describe the locus on which the symmetric matrix
$A_{ij} = y_{ij}$ has rank one, as
the matrix $(x_i x_j)$
has\footnote{
In general, for ${\mathbb P}^{n-1}$, so that there are $n$ homogeneous
coordinates $x_i$, it is straightforward to show that
the matrix $(x_i x_j)$ has only one nonvanishing eigenvalue,
given by
$\sum_i x_i^2$.  Alternatively, to see that it has rank one, note that
when multiplied into the vector $(a_1, a_2, \cdots, a_n)^T$, all
components of the result are proportional to $\sum_i a_i x_i$. 
} rank one.
Following \cite{Jockers:2012zr},
the PAXY model is an $O(1) = {\mathbb Z}_2$
gauge theory with fields $x_i$ and a superpotential of the form
\begin{displaymath}
W \: = \: p_{ij}\left( A(\phi)_{ij} - x_i x_j \right).
\end{displaymath}
(In the construction of \cite{Jockers:2012zr}, the $x_i$ are merely 
introduced to define the locus on which $A(\phi)$ has rank one; in the
present context, however, they are homogeneous coordinates on the
source projective space of the Veronese embedding.)
Here, for $A(\phi)_{ij} = y_{ij}$, we have an identical superpotential.
Furthermore, because the $y_{ij}$ are charge two and the $x_i$ charge one
under the gauged $U(1)$, that means the $y_{ij}$ are
neutral under ${\mathbb Z}_2 \subset U(1)$, whereas the $x_i$ receive
a sign flip, also matching \cite{Jockers:2012zr}.

In hindsight, the fact that the GLSM for the degree two Veronese model
can be described as a PAXY model should not be a surprise, as the
image of degree two Veronese embeddings are Pfaffian varieties 
(see {\it e.g.} \cite{harris}[lecture 2]), 
hence
describable by PAX and PAXY models.

Analogous descriptions extend to the images of
Veronese embeddings of higher degree.  For a degree $d$ map,
label the homogeneous coordinates on the target projective space
by $y_{i_1 \cdots i_d}$, symmetric in its indices, then labelled the
Lagrange multipliers by $p_{i_1 \cdots i_d}$ similarly, we can write the
superpotential as
\begin{displaymath}
W \: = \: p_{i_1 \cdots i_d} \left( y_{i_1 \cdots i_d} - x_{i_1}
x_{i_2} \cdots x_{i_d} \right).
\end{displaymath}
Just as there was a gauged ${\mathbb Z}_2$ in the degree two case,
matching \cite{Jockers:2012zr}, here there is analogously a gauged
${\mathbb Z}_d$, corresponding to a subgroup of the $U(1)$ gauge
symmetry under which the $y_{i_1 \cdots i_d}$ are neutral but the
$x_i$ are charged.  This is an analogue of the PAXY model for a 
`hyperdeterminantal' variety, as one would expect for the image of a
Veronese map of higher degree.

In passing, note that the nonminimal charges of the $y$'s, dictated by gauge
invariance of the superpotential, also reflect the degree of the Veronese
embedding.  For a degree $d$ Veronese embedding, 
all the maps of the sigma model into the target projective space have
degree divisible by $d$.  
Sigma models with restrictions
on degrees of maps have been discussed in {\it e.g.} \cite{Pantev:2005zs,Pantev:2005rh,Hellerman:2006zs,Sharpe:2014tca} and references therein,
and in those references, indeed, GLSMs for such sigma models are described
by nonminimal $U(1)$ charges, divisible by a factor corresponding to the
divisibility properties of the maps, in exactly the form we see here.

Let us analyze the model for the degree $d$ Veronese embedding
more systematically, following the same pattern as discussed
for the PAXY model in \cite{Jockers:2012zr}[section 3.3].  
First, D-terms describe the space of $y$'s and $x$'s
as a weighted projective space ${\mathbb P}_{[1,\cdots,1,d,\cdots,d]}$,
of dimension
\begin{displaymath}
n \: + \: \left( \begin{array}{c} n + d\\ d \end{array} \right).
\end{displaymath}
The $p$'s then form a vector bundle over this projective space, 
defined by the sum of
\begin{displaymath}
\left( \begin{array}{c} n + d \\ d \end{array} \right)
\end{displaymath}
copies of ${\cal O}(-d)$.  In this language, the GLSM is realizing
a complete intersection in the weighted projective space
${\mathbb P}_{[1,\cdots,1,d,\cdots,d]}$.
For the same reasons discussed earlier, this complete intersection
can never intersect the ${\mathbb Z}_d$ singularities in that weighted
projective space.  Any singularities would have to arise in the
complete intersection itself, not from the ambient weighted projective
space.  However, there are no singularities in the complete intersection.
Defining
\begin{displaymath}
E_{i_1 \cdots i_d} \: = \: y_{i_1 \cdots i_d} - x_{i_1} \cdots x_{i_d},
\end{displaymath}
the derivatives
\begin{displaymath}
\left.
\frac{ \partial E_{i_1 \cdots i_d} }{
\partial (x, y ) } \right|_X
\end{displaymath}
form a set of sections of the normal bundle to the locus $X \equiv \{ E_{i_1 \cdots i_d}=0\}$,
and their linear independence guarantees that the
F-term conditions
\begin{displaymath}
p_{i_1 \cdots i_d} \frac{ \partial E_{i_1 \cdots i_d} }{
\partial (x, y ) }
\: = \: 0
\end{displaymath}
can only be solved by taking all the $p_{i_1 \cdots i_d} = 0$.
Thus, the given GLSM represents the Veronese embedding of ${\mathbb P}^n$,
as advertised.

As another consistency check, we verify below that the
Calabi-Yau condition realized in such GLSMs by demanding that the
one-loop correction to the Fayet-Iliopoulos parameter vanish,
coincides with the mathematical condition to have a Calabi-Yau.

To see this, note that the sum of the $y_i$ charges above
cancels the sum of the $p_i$ charges above, so that the sum of the
charges in a GLSM realizing just the Veronese embedding is the
same as the sum of the charges in the original projective space,
the sum of the charges of the $x_a$'s.  Therefore, in order for the
FI to be RG-invariant, one must take a hypersurface (or complete
intersection) whose $U(1)$ charge matches the sum of the charges
of the $x_a$'s.  This is the same as the same as the Calabi-Yau 
condition.

To be concrete, consider the example of a degree three Veronese
embedding ${\mathbb P}^2 \rightarrow {\mathbb P}^9$:
\begin{displaymath}
[x_0,x_1,x_2] \: \mapsto \: [x_0^3, x_1^3, x_2^3,
x_1^2 x_2, x_1^2 x_3, x_2^2 x_1, x_2^2 x_3, x_3^2 x_1, x_3^2 x_2,
x_1 x_2 x_3].
\end{displaymath}
To get a Calabi-Yau, we intersect the image with a hyperplane
(a hypersurface of degree one) in ${\mathbb P}^9$.
This may be counterintuitive, as when dealing with complete
intersections, we work with the sum of the degrees, but this is not
a complete intersection.  To see more explicitly that the result is
Calabi-Yau, let us consider an example of a degree one hypersurface
in ${\mathbb P}^9$ intersecting the image of the embedding:
\begin{displaymath}
y_0 + y_1 + y_2 \: = \: x_0^3 + x_1^3 + x_2^3.
\end{displaymath}
Clearly, the intersection of
a hyperplane in ${\mathbb P}^9$ with the image of the (degree 3)
Veronese embedding
is equivalent to a cubic
in ${\mathbb P}^2$, which is Calabi-Yau.  

To realize the example above in a GLSM, we consider a theory with
gauge group $U(1)$, and matter
\begin{itemize}
\item 3 chiral superfields $x_i$ of charge $1$, corresponding to homogeneous
coordinates on the original ${\mathbb P}^2$,
\item 10 chiral superfields $y_j$ of charge $3$, corresponding to homogeneous
coordinates on the image ${\mathbb P}^9$,
\item 10 chiral superfields $p_j$, charge $-3$, acting as Lagrange
multipliers identifying the homogeneous coordinates of the ${\mathbb P}^9$
with the image of the Veronese embedding,
\item one chiral superfield $q$ of charge $-3$, realizing a hyperplane
in ${\mathbb P}^9$,
\end{itemize}
and superpotential
\begin{displaymath}
W \: = \: \sum_{j=0}^9 p_j\left( y_j - \nu_j(x) \right) \: + \:
q G(y),
\end{displaymath}
where $\nu_j(x)$ is the $j$th entry in the Veronese embedding
({\it e.g.} $\nu_0(x) = x_0^3$, $\nu_1(x) = x_1^3$, and so forth),
and $G(y)$ is linear in $y$'s (hence gauge charge $3$).

The F terms imply
\begin{displaymath}
G(y) = 0, \: \: \: 
y_j = \nu_j(x), \: \: \:
\sum_j p_j \partial_i \nu_j(x) = 0, \: \: \:
p_j +  q \frac{\partial G}{\partial y_j} = 0.
\end{displaymath}
At any fixed (nonzero) vev of the $x$'s (and hence $y$'s), the last
two equations form 13 linear homogeneous
equations for 11 unknowns (the $p$'s and $q$), hence generically
one expects that $q=0$ and all the $p_j=0$, as expected for the
geometric interpretation we have described.

Note that the sum of the $U(1)$ charges is zero -- the charges of the
$p_j$'s and $y_j$'s cancel out, leaving one just with the
charges of the $x_i$'s and $q$.  Thus, as expected, we see that our
GLSM description of the Veronese embedding correctly realizes the
Calabi-Yau condition, as the condition that the Fayet-Iliopoulos parameter
does not run and so is RG-invariant.

Let us also compare the structure of mathematical singularities with
physical singularities.  One way that the intersection of a Veronese embedding 
with a hypersurface can be singular is if the hypersurface is singular (and
the Veronese embedding intersects the image).  Briefly, if one adds a 
term $q G(y)$, corresponding to a hypersurface $\{G(y)=0\}$,
and the hypersurface is singular, then for some point $y$,
both $G(y)=0$ as well as the derivatives of $y$.  If that point intersects
the image of the Veronese embedding, then for the corresponding $y$ vevs,
F term constraints on $q$ drop out, and so the theory develops an
unbounded branch of $q$ vevs, intersecting the original geometry
at the singular point.

There are other ways in which an intersection of a Veronese embedding
with a hypersurface can develop a singularity.
For example, let us return to the case of a hyperplane in our previous
example.
Suppose that the hyperplane $\{ G(y) = 0 \} \subset {\mathbb P}^9$ is
defined by $G(y)$ containing a single monomial, such that the
intersection with the image of ${\mathbb P}^2$ is $x_0^2 x_1$.
Mathematically,
although a generic hyperplane in ${\mathbb P}^9$ is smooth, this intersection
is singular, at the point $x_0 = 0 = x_1$.  Physically, we can see
this singularity as follows.  

To be specific, suppose $G(y) = y_n$ for some homogeneous coordinate
$y_n$.  The F term constraints are
\begin{displaymath}
y_n = 0, \: \: \:
y_j = \nu_j(x), \: \: \:
\sum_j p_j \partial_i \nu_j(x) = 0, \: \: \:
p_n + q = 0, \: \: \:
p_i = 0 \mbox{ for } i \neq n.
\end{displaymath}
We see immediately that most of the $p$'s vanish, except possible
for $p_n$, which is constrained to equal $-q$.
The F term constraints arising from the $x$'s then have the form
\begin{displaymath}
p_n \partial_i \nu_n(x) = -q \partial_i (x_0^2 x_1 ) = 0,
\end{displaymath}
hence
\begin{displaymath}
-q (2 x_0 x_1) = 0 = -q (x_0^2).
\end{displaymath}
At the singular point $x_0 = x_1 = 0$ ($x_2 \neq 0$),
$q$ and hence $p_n$ have unbounded magnitudes, hence we have a noncompact
branch, indicating a singularity in the GLSM.

So far we have discussed intersections of the image of the Veronese
embedding with hypersurfaces in the target space.  In principle, one can
also discuss images of hypersurfaces under the Veronese embedding,
as well as their intersections with hypersurfaces in the target.
For example, let $f(x)$ define a hypersurface of degree $d_f$
in the (original)
projective space defined by the $x$'s, and $G(y)$ a hypersurface 
of degree $d_G$ in the target projective space, defined by the
$y$'s.  Suppose the $x$'s and $y$'s are related
by a Veronese embedding of degree $d$.  We could describe the intersection
of the image of $\{f(x)=0\}$ with $\{ G(y)=0\}$ with a GLSM with gauge
group $U(1)$ and the following
fields:
\begin{itemize}
\item chiral superfields $x_i$ of charge $1$, corresponding to homogeneous
coordinates in the original projective space,
\item chiral superfields $y_j$ of charge $d$, corresponding to homogeneous
coordinates in the target projective space,
\item chiral superfields $p_j$ of charge $-d$ enforcing the Veronese
map,
\item a chiral superfield $s$ of charge $-d_f$,
\item a chiral superfield $q$ of charge $-d d_G$,
\end{itemize}
and superpotential
\begin{displaymath}
W \: = \: \sum_j p_j (y_j - \nu_j(x)) \: + \:
s f(x) \: + \: q G(y).
\end{displaymath}
The F-terms imply, for example,
\begin{displaymath}
y_j = \nu_j(x), \: \: \:
f(x) = 0, \: \: \: G(y) = 0,
\end{displaymath}
and in this fashion one
sees that, in the $r \gg 0$ phase, this GLSM is describing the
intersection of the image of $\{f(x)=0\}$ with $\{G(y)=0\}$.

For completeness, let us also consider the $r \ll 0$ limit of these
models.  To be concrete, consider
a quadratic Veronese embedding, say ${\mathbb P}^1
\rightarrow {\mathbb P}^2$.  Here we have
\begin{itemize}
\item two chiral superfields $x_i$ of charge $1$,
\item three chiral superfields $y_j$ of charge $2$,
\item three chiral superfields $p_j$ of charge $-2$,
\end{itemize}
and superpotential
\begin{displaymath}
W \: = \: p_0( y_0 - x_0^2) \: + \: p_1( y_1 - x_1^2) \: + \:
p_2(y_2 - x_0 x_1) .
\end{displaymath}
For $r \ll 0$, D terms require that the $p$'s are not all zero.
At the same time, the F terms for the $y_j$ require that all the $p$'s vanish.
This is a contradiction -- classically, there are no\footnote{
In passing, note that if there were supersymmetric vacua, there might be
an interesting structure.  Write the superpotential as
\begin{displaymath}
W \: = \: p_j y_j \: - \: x_i A^{ij}(p) x_j,
\end{displaymath}
then $A^{ij}$ is a mass matrix for the $x$'s, given by
\begin{displaymath}
(A^{ij}) \: = \: \left[ \begin{array}{cc}
p_0 & 1/2 p_2 \\ 1/2 p_2 & p_1 \end{array} \right].
\end{displaymath}
Away from the locus $\{ \det A=0 \}$,
all massless fields have charge $\pm 2$, so we have a branched double
cover of Tot ${\cal O}(-1)^3 \rightarrow {\mathbb P}^2$, defined by
the $p$'s and $y$'s, branched over this degree two locus in ${\mathbb P}^2$,
with remaining superpotential terms $p_j y_j$
\cite{Caldararu:2007tc,Pantev:2005zs,Pantev:2005rh,Hellerman:2006zs}.
} supersymmetric
vacua for $r \ll 0$, just as in the ordinary 
${\mathbb P}^n$ model.

More generally, we can now build a GLSM describing the intersection of
the image of $k$ hypersurfaces in ${\mathbb P}^n$, under a degree $d$
Veronese embedding, with $m$ hypersurfaces in the target projective
space, of dimension
\begin{displaymath}
\left( \begin{array}{c}
n +d \\ d \end{array} \right) - 1.
\end{displaymath}
The expected dimension of this space will be $n-k-m$.
Suppose the $k$ hypersurfaces in ${\mathbb P}^n$ have degrees
$D_1, \cdots, D_k$, and the $m$ hypersurfaces in the target projective
space have degrees $E_1, \cdots, E_m$.  The Calabi-Yau condition
is
\begin{displaymath}
\left( \sum_i D_i \right) \: + \: d \left( \sum_j E_j \right) \: = \:
n+1.
\end{displaymath}
The corresponding GLSM is a $U(1)$ gauge theory with matter
\begin{itemize}
\item $n+1$ chiral superfields $x_i$ of charge $1$, corresponding to
homogeneous coordinates on ${\mathbb P}^n$,
\item $(n+d) \cdots (n+1) / d!$ chiral superfields $y_j$ of charge $d$,
corresponding to homogeneous coordinates on the target projective space,
\item $(n+d) \cdots (n+1) / d!$ chiral superfields $p_j$ of charge $-d$,
corresponding to Lagrange multipliers describing the Veronese embedding,
\item $k$ chiral superfields $q_a$ of charges $-D_1, \cdots, -D_k$,
\item $m$ chiral superfields $r_b$ of charges $-d E_1, \cdots, -d E_k$,
\end{itemize}
with superpotential
\begin{displaymath}
W \: = \: p_j \left( y_j - \nu_j(x) \right) \: + \:
q_a f_a(x) \: + \: r_b g_b(y),
\end{displaymath}
where $f_a(x)$, $g_b(y)$ are the homogeneous polynomials defining the
hypersurfaces in ${\mathbb P}^n$ and in the target projective space.
The Calabi-Yau condition follows immediately from the sum of the $U(1)$
charges.

The reader should note that the spaces above are not new to the literature --
these are simply the complete intersections
${\mathbb P}^n[D_1, \cdots, D_k, d E_1, \cdots, d E_m]$, described in an
unusual fashion.

For one example, we can obtain a Calabi-Yau threefold as the intersection
of the image of a hypersurface in ${\mathbb P}^5$ of degree $2$,
under a Veronese embedding of degree $2$,
with a hypersurface in the target projective space of degree $2$. 
This is equivalent to the complete intersection ${\mathbb P}^5[2,4]$.

There are several other closely analogous examples which are straightforward
to realize in GLSMs, which we will describe in subsequent subsections.
As the analysis is very similar to what we have just described,
our descriptions will be brief.

\subsection{Segre embedding}

The Segre embedding can be realized in a similar fashion to 
Veronese embeddings.  Recall the Segre map is the map
\begin{displaymath}
{\mathbb P}^k \times {\mathbb P}^n \: \longrightarrow \:
{\mathbb P}^{(k+1)(n+1)-1}
\end{displaymath}
as
\begin{displaymath}
[x_0, \cdots, x_k] \times [y_0, \cdots, y_n] \: \mapsto \:
[ x_0 y_0, x_0 y_1, \cdots, x_n y_n],
\end{displaymath}
where the $x_i$ are homogeneous coordinates on ${\mathbb P}^k$
and the $y_j$ are homogeneous coordinates on ${\mathbb P}^n$.

We can realize this map in a GLSM as follows.
Consider a $U(1) \times U(1)$ gauge theory, with matter
\begin{itemize}
\item $k+1$ chiral superfields $x_i$ of charge $(1,0)$,
\item $n+1$ chiral superfields $y_j$ of charge $(0,1)$,
\item $(k+1)(n+1)$ chiral superfields $z_a$ of charge $(1,1)$,
\item $(k+1)(n+1)$ chiral superfields $p_a$ of charge $(-1,-1)$,
acting as Lagrange multipliers forcing the $z_a$'s to be identified with the
image of the Segre embedding,
\end{itemize}
and superpotential
\begin{displaymath}
W \: = \: \sum_a p_a \left( z_a - S_a(x,y) \right),
\end{displaymath}
where $S_a(x,y)$ gives the image of the Segre embedding, meaning
$S_0 = x_0 y_0$, $S_1 = x_0 y_1$, and so forth.

The analysis of this GLSM follows much as before.  In the phase
defined by Fayet-Iliopoulos parameters $(r_1,r_2)$ where $r_1 \gg 0$ and
$r_2 \gg 0$, D-terms imply that not all the $x$'s vanish and not all the
$y$'s vanish, as one would expect for a pair of projective spaces.
F-terms require
\begin{displaymath}
z_a = S_a(x,y), \: \: \:
p_a = 0, \: \: \:
\sum_a p_a \frac{\partial S_a}{\partial x_i} = 0 = 
\sum_a p_a \frac{\partial S_a}{\partial y_j},
\end{displaymath}
so we immediately recover the desired geometry, as before.

For purposes of computing Calabi-Yau hypersurfaces in the image,
note that the charges of the $z_a$, $p_a$ cancel out in the renormalization
of the Fayet-Iliopoulos parameter, so that only the $x_i$, $y_j$ charges
contribute to the determination of the Calabi-Yau condition, in exactly
the same fashion as the Veronese embedding previously discussed.

As one simple example, the image of the Segre embedding
${\mathbb P}^1 \times {\mathbb P}^1 \rightarrow {\mathbb P}^3$
is a quadric hypersurface in ${\mathbb P}^3$, specifically the
hypersurface
$z_0 z_1 = z_2 z_3$, which can be seen by identifying
\begin{displaymath}
z_0 = x_0 y_0, \: \: \: z_1 = x_1 y_1, \: \: \:
z_2 = x_1 y_0, \: \: \: z_3 = x_0 y_1.
\end{displaymath}

We can slightly rewrite the GLSM above to emphasize another part of the
mathematics.  The image of a Segre embedding is a determinantal variety,
which admits PAX and PAXY descriptions, and the description above is
of the form of a PAXY model.  Specifically, write the target-space
homogeneous coordinates as $z_{ij}$, where the Segre embedding relates
$z_{ij} = x_i y_j$.
Then, the GLSM has a superpotential of the form
\begin{displaymath}
W \: = \: p_{ij} \left( z_{ij} - x_i y_j \right).
\end{displaymath}
Now, the image of the Segre embedding is the locus in
${\mathbb P}^{(k+1)(n+1)-1}$ given by the zero locus of the $2\times 2$ minors
of the
matrix with components $z_{ij}$.

Now, as will be reviewed more extensively in section~\ref{sect:int-pf:paxy},
the PAXY model \cite{Jockers:2012zr} 
that describes the locus in ${\mathbb P}^{(k+1)(n+1)-1}$
where the $N \times M$ matrix
$A(\phi)_{ij}$ has rank $m$ (corresponding to the vanishing
of all $(m+1)\times (m+1)$ minors)
is defined by a $U(m)$ gauge theory with 
\begin{itemize}
\item $N$ chiral superfields $x_{ia}$ in the fundamental of $U(m)$,
\item $M$ chiral superfields $y_i^a$ in the antifundamental of $U(m)$,
\item $N M$ chiral superfields $p_{ij}$ which are neutral under $U(m)$,
\end{itemize}
with superpotential
\begin{displaymath}
W \: = \: p_{ij}\left( A(\phi)_{ij} - x_{ia} y_j^a \right).
\end{displaymath}
Here, to describe the image of the Segre embedding, 
we wish to restrict to the rank-one locus of the
$(k+1)\times(n+1)$ matrix $z_{ij}$, and it is straightforward to check
that the matter content and superpotential of the Segre realization
match the PAXY model for this case, with the PAXY $U(1)$ corresponding to
the antidiagonal $U(1)$ of the Segre realization's $U(1) \times U(1)$.

Thus, as a consistency check, we see that the description of the Segre
embedding above duplicates the description of the image as a determinantal
variety via a PAXY description.

\subsection{Diagonal map}

The map
\begin{displaymath}
\Delta: \: {\mathbb P}^n \: \longrightarrow \:
{\mathbb P}^n \times {\mathbb P}^n
\end{displaymath}
embeds ${\mathbb P}^n$ diagonally into the product of projective
spaces.

The corresponding GLSM is a $U(1)$ gauge theory with matter
\begin{itemize}
\item $n+1$ chiral superfields $x_i$ of charge $1$, corresponding 
to homogeneous coordinates on the source ${\mathbb P}^n$,
\item $n+1$ chiral superfields $y_i$ of charge $1$, corresponding to
homogeneous coordinates on one of the target factors,
\item $n+1$ chiral superfields $z_i$ of charge $1$, corresponding to
homogeneous coordinates on the other factor,
\item $2(n+1)$ chiral superfields $p_i$, $q_i$ of charge $-1$,
\end{itemize}
and superpotential
\begin{displaymath}
W \: = \: p_i (y_i - x_i) \: + \:
q_i (z_i - x_i).
\end{displaymath}

Briefly, on the one hand, if we integrate out the $y$s and $z$s, we recover
the original projective space, whereas if we integrate out $x$s, we find the
superpotential
\begin{displaymath}
W \: = \: p'_i (y_i - z_i).
\end{displaymath}

In more detail, for $r \gg 0$, D-terms imply that not all the $x$s, $y$s,
and $z$s vanish.  F-terms then imply
\begin{displaymath}
x_i = y_i = z_i, \: \: \:
p_i = q_i = 0.
\end{displaymath}

Mathematically, the composition of the Segre and diagonal maps is equivalent
to the Veronese embedding of degree two.  
To see this, write ${\mathbb P}^n = {\mathbb P} V$ for a vector space
$V$ of dimension $n+1$.  Then
\begin{eqnarray*}
\Delta: & {\mathbb P} V & \longrightarrow \: 
{\mathbb P} V \times {\mathbb P} V,
\\
& z & \mapsto \: (z,z),
\end{eqnarray*}
and
\begin{eqnarray*}
{\rm Segre}: & {\mathbb P} V \times {\mathbb P} V & \longrightarrow \:
{\mathbb P} (V \otimes V), \\
& (x,y) & \mapsto x \otimes y.
\end{eqnarray*}
Thus, the composition is the map ${\mathbb P} V \rightarrow
{\mathbb P} (V \otimes V)$ given by $z \mapsto z \otimes z$.
However, this factors through ${\mathbb P}({\rm Sym}^2 V)$, so in
particular
\begin{displaymath}
{\rm Segre} \circ \Delta \: = \:
j \circ \nu_2,
\end{displaymath}
where $j: {\mathbb P}({\rm Sym}^2 V) \rightarrow
{\mathbb P}(V \otimes V)$.

Physically, we can represent the composition 
\begin{displaymath}
{\rm Segre} \circ \Delta: \:
{\mathbb P}^n \: \stackrel{\Delta}{\longrightarrow} \:
{\mathbb P}^n \times {\mathbb P}^n \: 
\stackrel{\rm Segre}{\longrightarrow} \: 
{\mathbb P}^{(n+1)^2-1}
\end{displaymath}
as follows.
Consider a GLSM with gauge group $U(1)$ and matter
\begin{itemize}
\item $n+1$ chiral superfields $x_i$ of charge $1$, corresponding to
homogeneous coordinates on the first source ${\mathbb P}^n$,
\item $n+1$ chiral superfields $y_i$ of charge $1$, corresponding to
homogeneous coordinates on the one of the middle ${\mathbb P}^n$ factors,
\item $n+1$ chiral superfields $z_i$ of charge $1$, corresponding to
homogeneous coordinates on another of the middle ${\mathbb P}^n$ factors,
\item $(n+1)^2$ chiral superfields $w_{ij}$ of charge $2$, corresponding
to homogeneous coordinates on the final ${\mathbb P}^{n^2+2n}$,
\item $2(n+1)$ chiral superfields $p_i$, $q_i$ of charge $-1$,
\item $(n+1)^2$ chiral superfields $s_{ij}$ of charge $-2$,
\end{itemize}
and superpotential
\begin{displaymath}
W \: = \: 
p_i(y_i - x_i) + q_i(z_i - x_i) +
s_{ij}(w_{ij} - y_i z_i).
\end{displaymath}
After integrating out the $p$'s, $q$'s, $y$'s, and $z$'s, this becomes
\begin{displaymath}
W \: = \: s_{ij}(w_{ij} - x_i x_j),
\end{displaymath}
which is explicitly the GLSM representing
$j \circ \nu_2: {\mathbb P}^n \rightarrow
{\mathbb P}^{n^2+2n}$.

\subsection{Nonabelian Veronese embedding}
\label{sec:nonab-ver}
There is an analogue of the Veronese embedding for Grassmannians,
see {\it e.g.} \cite{harris}[section 6.10],
which sends
\begin{displaymath}
\nu_d: \: G(k,n) \: \longrightarrow \: G\left( 
\left( \begin{array}{c} k+d \\ d \end{array} \right), 
\left( \begin{array}{c} n+d \\ d \end{array} \right) \right).
\end{displaymath}

We describe a conjecture\footnote{
As we will not use this example elsewhere, it has not been checked
as thoroughly as other cases discussed in this paper.
}
below for how this can be
realized in a GLSM. 
Consider a GLSM with gauge group $U(k)$, and matter as follows:
\begin{itemize}
\item $n$ chiral superfields $\Phi_i^a$ in the representation
${\bf k}$ of the gauge group,
\item $\left( \begin{array}{c} n + d \\ d \end{array} \right)$ chiral
superfields $\tilde{\Phi}_{i_1 \cdots i_d}^{a_1 \cdots a_d}$,
$i_1 \leq \cdots \leq i_d$, $a_1 \leq \cdots \leq a_d$
(symmetric in both sets of indices),
in the representation ${\bf {\rm Sym}^d k}$
of the gauge group,
\item $\left( \begin{array}{c} n + d \\ d \end{array} \right)$
chiral superfields $p^{i_1 \cdots i_d}_{a_1 \cdots a_d}$, 
$i_1 \leq \cdots \leq i_d$, $a_1 \leq \cdots \leq a_d$ (symmetric in
both sets of indices)
in the representation ${\bf {\rm Sym}^d \overline{k}}$ 
of the
gauge group,
\end{itemize}
with superpotential
\begin{displaymath}
W \: = \: \sum_{i_1 \leq \cdots \leq i_d} p^{i_1 \cdots i_d}_{a_1 \cdots a_d} 
\left(
\tilde{\Phi}_{i_1 \cdots i_d}^{a_1 \cdots a_d} - 
{\rm Sym} \,\Phi_{i_1}^{a_1} \cdots \Phi_{i_d}^{a_d} \right).
\end{displaymath}
The analysis of this model is much the same as in previous sections.
Note for the purpose of defining Calabi-Yau hypersurfaces by the RG
running of the Fayet-Iliopoulos parameter that the contributions from the
$p$'s and $\tilde{\Phi}$'s cancel out, so that this behaves just like an
ordinary Grassmannian $G(k,n)$, and furthermore that integrating out
$\tilde{\Phi}$'s yields the ordinary Grassmannian model, as expected.

One point of confusion is that in the image Grassmannian,
we only take the symplectic quotient by $U(k)$ not
\begin{displaymath}
U\left( \begin{array}{c} k + d \\ d \end{array} \right).
\end{displaymath}
We have not checked carefully, but we do observe that this appears to be
analogous to the fact in earlier Veronese embeddings that one quotiented
by a different gauge group (a cover of $U(1)$ or of $U(k)$), corresponding
to the fact that maps into the target space all had certain divisibility
properties, as in {\it e.g.} 
\cite{Pantev:2005zs,Pantev:2005rh,Hellerman:2006zs,Sharpe:2014tca}.
Here, the analogous divisibility properties would be rather more complicated.

\subsection{More general maps}    \label{sect:general-maps}

Given any homogeneous map $f: {\mathbb P}^n \rightarrow {\mathbb P}^m$,
by which we mean a set of $m+1$ homogeneous polynomials, we can similarly
define a corresponding GLSM.

Suppose the components of the map $f$ all have degree $k$.
Then, the corresponding GLSM is a $U(1)$ gauge theory with
\begin{itemize}
\item $n+1$ chiral multiplets $x_i$ of charge $1$,
\item $m+1$ chiral multiplets $y_j$ of charge $k$,
\item $m+1$ chiral multiplets $p_j$ of charge $-k$,
\end{itemize}
and superpotential
\begin{displaymath}
W \: = \: p_j( y_j - f_j(x) ).
\end{displaymath}

Strictly speaking, this GLSM describes the graph of the function $f(x)$,
which for a map $f: X \rightarrow Y$ is defined to be 
\begin{displaymath}
\{ (x,y) \in X \times Y \, | \, y = f(x) \}.
\end{displaymath}
The graph of $f$ always embeds into $X \times Y$,
so the graph is always isomorphic to $X$, mathematically.
Physically, the same result follows from the fact that $p$'s and $y$'s can
be integrated out.  The graph will be isomorphic to the image of $f$
if and only if $f$ is an embedding.

For a trivial example, consider a constant map $f$ on ${\mathbb P}^n$.
After absorbing the constant into a trivial field redefinition, the
superpotential can be written
\begin{displaymath}
W \: = \: p_j (y_j - 0).
\end{displaymath}
The F terms require that $p$ and $y$ vanish, leaving one just with a
copy of the supersymmetric ${\mathbb P}^n$ model, matching the
mathematical prediction.

In the cases discussed so far (Veronese, Segre, diagonal embeddings),
because they are embeddings,
the image of the map is isomorphic to the source, and so the graph is
isomorphic to the image.

For completeness, let us briefly outline composition.
Given a map $f: {\mathbb P}^n \rightarrow {\mathbb P}^m$ of degree $k$,
and a map $g: {\mathbb P}^m \rightarrow {\mathbb P}^r$ of degree $\ell$,
we take the GLSM corresponding to the composition to be
a $U(1)$ gauge theory with matter
\begin{itemize}
\item $n+1$ chiral multiplets $x_i$ of charge $1$,
\item $m+1$ chiral multiplets $y_j$ of charge $k$,
\item $r+1$ chiral multiplets $z_a$ of charge $k \ell$,
\item $m+1$ chiral multiplets $p_j$ of charge $-k$,
\item $r+1$ chiral multiplets $q_a$ of charge $-k \ell$,
\end{itemize}
and superpotential
\begin{displaymath}
W \: = \:
p_j(y_j - f_j(x) ) \: + \:
q_a(z_a - g_a(y) ).
\end{displaymath}

Now, in the expression above, we can trivially integrate out $y$'s and $p$'s,
to get a $U(1)$ gauge theory of $x$'s, $z$'s, and $q$'s, with superpotential
\begin{displaymath}
W \: = \: q_a \left(z_a - g_a(f(x)) \right),
\end{displaymath}
trivially corresponding to the map $g \circ f$.  For this reason, we can
identify the GLSM above with the composition of the GLSM's for $f$ and
$g$.

\subsection{Another perspective on hypersurfaces}

As an entertaining aside, we can describe the GLSM for an ordinary
hypersurface in a projective space in the language of maps.

Specifically, first interpret the defining equation of a hypersurface,
a homogeneous polynomial $f$ of degree $d$, 
as a map $f: {\mathbb P}^n \rightarrow
{\mathbb C}$.
Then, take the intersection with the origin of ${\mathbb C}$,
effectively realizing $f^{-1}(0)$.  

We can describe this in a GLSM as
follows.  Consider a $U(1)$ gauge theory with matter
\begin{itemize}
\item $n+1$ chiral superfields $x_i$ of charge $1$, corresponding to
homogeneous coordinates on ${\mathbb P}^n$,
\item one chiral superfield $y$ of charge $d$, corresponding to the
image of $f$,
\item two chiral superfields $p$, $q$ of charge $d$,
\end{itemize}
and superpotential
\begin{displaymath}
W \: = \: q(y - f(x)) \: + \: p( y - 0 ).
\end{displaymath}
Integrating out $y$ and $q$, one recovers the standard GLSM for a hypersurface.

\section{General aspects of intersections of Grassmannians and Pfaffians}
\label{sect:int-nonabelian}

\subsection{Warm-up:  Complete intersections of hypersurfaces}

This section will be concerned with novel GLSM descriptions of complete
intersections of Grassmannians and Pfaffians,
so as a warm-up, we will describe a pertinent presentation of a complete
intersection of hypersurfaces in a projective space, to illustrate a basic
idea we will use later.

Consider hypersurfaces in ${\mathbb P}^n$ defined by two homogeneous
polynomials $G_1(\phi)$, $G_2(\phi)$ of degrees $d_1$, $d_2$, respectively.  Ordinarily, we would describe
this by a GLSM with $n+1$ chiral superfields $\phi_i$ of charge $1$
and two additional chiral superfields $p_1$, $p_2$ of charges $-d_1$,
$-d_2$, respectively, with superpotential 
\begin{displaymath}
W \: = \: p_1 G_1(\phi) \: + \: p_2 G_2(\phi).
\end{displaymath}

However, there is an alternative presentation that one could use instead.
Consider an alternative theory with $2(n+1)$ chiral superfields $\varphi_i$,
$\tilde{\varphi}_i$ ($i \in \{1, \cdots, n+1\}$)
of charge $1$, two chiral superfields $p_1$, $p_2$ of charges
$-d_1$, $-d_2$, respectively, as before, plus $n+1$ chiral superfields
$b_i$ of charge $-1$,
with superpotential
\begin{displaymath}
W \: = \: p_1 G_1(\varphi) \: + \: p_2 G_2(\tilde{\varphi}) \: + \:
\sum_{i=1}^{n+1} b_i (\varphi_i - \tilde{\varphi}_i).
\end{displaymath}
In this case, $p_1$ acts as a Lagrange multiplier corresponding to the
hypersurface $\{ G_1 = 0 \}$ in the space of $\varphi$'s,
$p_2$ acts as a Lagrange multiplier corresponding to the hypersurface
$\{ G_2 = 0 \}$ in the space of $\tilde{\varphi}$'s,
and the $b$'s act as Lagrange multipliers identifying homogeneous
coordinates on the two copies of ${\mathbb P}^{n}$.

Let us check this more thoroughly.  For $r \gg 0$, D terms require that
$\{ \varphi_1, \cdots, \tilde{\varphi}_{n+1} \}$
are not all simultaneously zero.
F terms require
\begin{displaymath}
G_1 = 0, \: \: \:
G_2 = 0, \: \: \:
\varphi_i - \tilde{\varphi}_i = 0, \: \: \:
p_1 \partial_i G_1 + b_i  = 0, \: \: \:
p_2 \partial_i G_2 - b_i = 0.
\end{displaymath}
which requires
\begin{displaymath}
G_1 = 0, \: \: \:
G_2 = 0, \: \: \:
\varphi_i - \tilde{\varphi}_i = 0, \: \: \:
p_1 \partial_i G_1(\varphi) + p_2 \partial_i G_2(\tilde{\varphi}) = 0,
\end{displaymath}
which after collapsing $\tilde{\varphi} \rightarrow \varphi$ reduce
to the same conditions as in ordinary presentations of
complete intersections.  Nondegeneracy of the complete intersection
requires $p_1 = p_2 = 0$ and so $b_i=0$.

This second description is overkill for the simple case of a complete
intersection of hypersurfaces in a projective space, but will be a prototype
for intersections of Grassmannians and Pfaffian varieties.

\subsection{Intersections of Grassmannians}
\label{sec:int-grass}
Next, consider a Grassmannian $G(k,N)$.  This is described \cite{Witten:1993xi}
by a $U(k)$ gauge theory with $N$ chiral multiplets 
$\phi^i_a$ in the fundamental
representation.  It has a natural embedding into a projective space
of dimension 
\begin{displaymath}
\left( \begin{array}{c}
N \\ k \end{array} \right) \: - \: 1,
\end{displaymath}
defined by the $SU(k)$-invariant combinations of chiral superfields
\begin{displaymath}
B^{i_1 \cdots i_k} \: = \: 
\phi^{i_1}_{a_1} \cdots \phi^{i_k}_{a_k} \epsilon^{a_1 \cdots a_k}.
\end{displaymath}
These act as homogeneous coordinates on the projective space,
and define what is known mathematically as the Pl\"ucker embedding.

\subsubsection{Self-intersection with linear rotation}
\label{sec:int-grass-rot}
Given a Grassmannian $G(k,N)$, one can deform the Plucker embedding by
replacing the baryons with linear combinations, and then intersect the
deformed image with the original Pl\"ucker embedding.  
This is what we refer to as a `self-intersection of Grassmannians,'
the intersection of $G(k,N)$ with a linear deformation in its Pl\"ucker
embedding.  We will restrict to the case that $N$ odd in the following.

The expected dimension of this intersection is\footnote{
First note that the image of $G(k,N)$ in its Pl\"ucker
embedding is locally described by
\begin{displaymath}
\left( \begin{array}{c} N \\ k \end{array} \right) - 1 - k(N-k)
\end{displaymath}
equations.  The expected dimension of the intersection is the dimension of the
ambient space minus the number of equations, which for a general linear
deformation, will be twice the number of equations above, hence the
expected dimension is
\begin{displaymath}
\left( \begin{array}{c} N \\ k \end{array} \right) - 1
 - 2 \left[ \left( \begin{array}{c} N \\ k \end{array} \right) - 1 - k(N-k)
\right] \: = \:
2 k(N-k) + 1 - \left( \begin{array}{c} N \\ k \end{array} \right).
\end{displaymath}
}
\begin{equation}  \label{eq:int-grass:dim}
2 k(N-k) + 1 - \left( \begin{array}{c} N \\ k \end{array} \right).
\end{equation}
Furthermore, if the expected dimension is negative, then the intersection
will be empty, so to have a nonempty intersection, one must require
\begin{equation}  \label{eq:grass-nonempty}
2 k(N-k) + 1 \: \geq \:
\left( \begin{array}{c} N \\ k \end{array} \right).
\end{equation}
As a result, these intersections will typically be of interest only
for relatively small values of $N$ and $k$.

The intersection will be smooth when the two Grassmannians intersect
transversely, which will happen (for positive expected dimension) for
generic rotations of one copy.  Note however that for trivial rotations,
equal to the identity, the dimension of the intersection jumps,
so that the family is not flat at this point.
(We will see this reflected in the occurrence of noncompact branches in
the GLSM at such points.)

This intersection will be Calabi-Yau when
\begin{equation}   \label{eq:int-grass:cy}
\left( \begin{array}{c} N \\ k \end{array} \right) \: = \:  2 N.
\end{equation}
We can see this as follows.  Let ${\mathbb P}$ denote the projective space
in which each copy of $G(k,N)$ is embedded, and $X$ the intersection of
the two copies.  For either Grassmannian $G(k,N)$, the adjunction formula
says that
\begin{displaymath}
K_G \: = \: K_{\mathbb P} |_G  + c_1\left( 
N_{G/{\mathbb P}}
\right),
\end{displaymath}
where $G$ denotes the Grassmannian, and for any $X$, $K_X$ denotes $c_1$
of the canonical bundle of $X$.  Now, $TG = S^* \otimes Q$, for
$S$ the universal subbundle (rank $k$) 
and $Q$ the universal quotient bundle (rank $N-k$).
Since $c_1(S) = {\cal O}(-1)$, $c_1(Q) = {\cal O}(+1)$, we deduce
$c_1(TG) = (N-k) c_1(S^*) + k c_1(Q) = {\cal O}(N)$.  We then have
\begin{displaymath}
K_G = -N, \: \: \:
K_{\mathbb P} = 
- \left( \begin{array}{c} N \\ k \end{array} \right),
\end{displaymath}
hence
\begin{displaymath}
c_1\left( N_{G/{\mathbb P}} \right) \: = \:
\left( \begin{array}{c} N \\ k \end{array} \right) - N.
\end{displaymath}
If the intersection $X$ is transversal, then
$N_{X/G_1} \cong N_{G_2/{\mathbb P}} |_X$, where
$G_1$ and $G_2$ denotes the two Grassmannians in ${\mathbb P}$.
From the adjunction formula,
\begin{displaymath}
K_X \: = \: K_{G_1}|_X  +
c_1\left( N_{X/G_1} \right) \: = \:
K_{G_1}|_X  +
c_1\left( N_{G_2/{\mathbb P}} |_X \right) \: = \:
\left( \begin{array}{c} N \\ k \end{array} \right) - 2N.
\end{displaymath}
Hence, we recover condition~(\ref{eq:int-grass:cy}) for the intersection
to be Calabi-Yau.

As an aside, it is straightforward to check that only in the case
$k=2$, $N=5$ will this intersection be Calabi-Yau.

We can describe
this intersection of $G(k,N)$ with itself in a GLSM as follows.  Take a 
\begin{displaymath}
\frac{
U(1) \times SU(k) \times SU(k)
}{
{\mathbb Z}_k \times {\mathbb Z}_k
}
\end{displaymath}
gauge theory, where each ${\mathbb Z}_k$ acts as the center of one $SU(k)$
factor and also on the $U(1)$,
with $N$ chiral multiplets $\phi^i_a$ in the
${\bf (k,1)_1}$ representation of the gauge group,
$N$ chiral multiplets $\tilde{\phi}^i_a$ in the
${\bf (1,k)_1}$ representation of the gauge group,
and
\begin{displaymath}
\left( \begin{array}{c}
N \\ k \end{array} \right)
\end{displaymath}
chiral superfields $p_{i_1 \cdots i_k}$ (antisymmetric in their indices)
in the ${\bf (1,1)_{-k}}$ representation,
with superpotential
\begin{displaymath}
W \: = \: p_{i_1 \cdots i_k} \left( f^{i_1 \cdots i_k}(B) -
\tilde{B}^{i_1 \cdots i_k} \right),
\end{displaymath}
where 
\begin{displaymath}
B^{i_1 \cdots i_k} \: = \: \phi^{i_1}_{a_1} \cdots \phi^{i_k}_{a_k}
\epsilon^{a_1 \cdots a_k}, \: \: \:
\tilde{B}^{i_1 \cdots i_k} \: = \: 
\tilde{\phi}^{i_1}_{a'_1} \cdots \tilde{\phi}^{i_k}_{a'_k}
\epsilon^{a'_1 \cdots a'_k},
\end{displaymath}
are the baryons corresponding to two copies of the Grassmannian, and
the $f^{i_1 \cdots i_k}$ (antisymmetric in indices) are linear functions
describing the deformation of one copy of the Grassmannian.
We claim that, at least for some values of $n$ and $k$, the GLSM above
will describe the intersection of the Grassmannians.

As one quick consistency check, it is straightforward to check that the
sum of the $U(1)$ charges will vanish when condition~(\ref{eq:int-grass:cy})
is satisfied, so we recover the Calabi-Yau condition as expected.

Let us discuss the discrete quotient in some more detail. Take
$\alpha\in U(1)$ and $g_1,g_2\in SU(k)$. Then we have the explicit
transformation rules
\begin{displaymath}
  p_{i_1i_2}\rightarrow\alpha^{-k}p_{i_1i_2}\qquad
  \phi^i\rightarrow\alpha g_1\phi^i\qquad
  \tilde{\phi}^i\rightarrow\alpha g_2\tilde{\phi}^i
  \end{displaymath}
We see there is a $\mathbb{Z}_k$ fixed locus at
\begin{displaymath}
  \alpha=e^{\frac{2\pi i}{k}j}\qquad g_1=g_2=e^{-\frac{2\pi i
      }{k}j}{\bf 1}, \qquad j=0,\ldots,k-1.
\end{displaymath}
Modding out by this $\mathbb{Z}_k$ action we can rewrite the gauge
group as $G=U(k)_{\alpha}\times U(k)_{\alpha}$ where by
$U(k)_{\alpha}$ we mean that the two $U(k)$s have the same central
U(1). Explicitly, this is realized by defining $\tilde{g}_1=\alpha
g_1\in U(k)$ and $\tilde{g}_2=\alpha g_2\in U(k)$. Note in particular
that $\det \tilde{g}_1=\alpha^k\det g_1=\alpha^k=\det\tilde{g}_2$,
which implies that $p_{i_1i_2}$ transforms in the inverse
determinantal representation of each of the $U(k)$s whereas the
baryons transform in the determinantal representation, and
hence the Calabi-Yau condition is satisfied.

Following \cite{Hori:2006dk} we can write down the D-terms associated
to the two $U(k)$s:
\begin{eqnarray*}
  U(k)_{\alpha,1}:&\quad& \sum_{i=1}^N \phi^i_a\phi^{i\dagger,b} -
  \sum_{j=1}^{N\choose k}|p_j|^2\delta_a^b
  -r\delta_a^b\\ 
U(k)_{\alpha,2}: &\quad&
  \sum_{i=1}^N \tilde{\phi}^i_{a'}\tilde{\phi}^{i\dagger,b'} -
  \sum_{j=1}^{N \choose k}|p_j|^2\delta_{a'}^{b'}-r\delta_{a'}^{b'},
  \end{eqnarray*}
where $r$ is the FI-parameter (for the single overall
$U(1)$) and $j$ combines the
$N\choose k$ independent
  index-combinations of the $p$-fields. For $r\gg0$ this implies that
  the the matrices $\phi\phi^{\dagger}$ and
  $\tilde{\phi}\tilde{\phi}^{\dagger}$ must have full rank, implying
that the $\phi^i_a$ vevs can be taken to form an orthonormal set of
vectors in ${\mathbb C}^N$ \cite{Witten:1993xi}[section 4.2],
whereas
  for $r\ll0$ not all $p$-fields are allowed to vanish
  simultaneously.

It is also interesting to add and subtract the two
  sets of D-terms. In the former case one gets
\begin{displaymath}
  \sum_{i=1}^N\phi^i_a\phi^{i\dagger,b} +
  \sum_{i=1}^N\tilde{\phi}^i_{a}\tilde{\phi}^{i\dagger,b}
  -2\sum_{k=1}^{N \choose k}|p_k|^2\delta_a^b-2\:r\delta_a^b.
  \end{displaymath}

The diagonal terms correspond to the $U(1)$ D-term. The difference of
the two D-terms gives
\begin{displaymath}
 \sum_{i=1}^N\phi^i_a\phi^{i\dagger,b} -
 \sum_{i=1}^N\tilde{\phi}^i_a\tilde{\phi}^{i\dagger,b}.
\end{displaymath}

For later reference let us also give an explicit parametrization of
the maximal torus of the gauge group. We first start off with
$U(1)\times SU(k)\times SU(k)$. We denote $\sigma_0$ for $U(1)$ and
$\mathrm{diag}(\sigma_{h,1},\ldots,\sigma_{h,k})$ ($h=1,2$) for
$SU(k)$, where we have $\sum_{j=1}^k\sigma_{h,j}=0$. The elements for
the two $U(k)_{\alpha}$ are
$\mathrm{diag}(\sigma_0+\sigma_{h,1},\ldots,\sigma_0+\sigma_{h,k})$. It is
convenient to make the following change of coordinates.
\begin{displaymath}
  \alpha_1=\sigma_0+\sigma_{1,1}\quad\ldots\quad
  \alpha_k=\sigma_0+\sigma_{1,k}\qquad
  \beta_1=\sigma_0+\sigma_{2,1}\quad\ldots\quad
  \beta_k=\sigma_0+\sigma_{2,k}
\end{displaymath}
These coordinates are subject to the condition
$\sum_{j=1}^k\alpha_k=\sum_{j=1}^j\beta_k$. We will use this to
explicitly eliminate one of them. Further note the positive roots of
$SU(k)$ are ${\bf e}_i-{\bf e}_{i+1}$ where the ${\bf e}_j$ are the
basis vectors of $\mathbb{R}^k$.

This information enters the effective potential on the Coulomb branch. The loci where the GLSM develops a Coulomb branch correspond to the singularities in the K\"ahler moduli space. Since our example is a one-parameter model, there are no mixed Coulomb-Higgs branches. The effective potential is
\begin{displaymath}
  \widetilde{W}_{eff}=-t(\sigma)
  -\sum_{i}Q_i(\sigma)(\log(Q_i(\sigma))-1) +\pi
  i\sum_{\rho>0}\rho(\sigma),
\end{displaymath}
where $t=r-i\theta$ is the FI-theta parameter, $Q_i$ are the gauge charges of the matter fields, $\sigma$ parametrizes the maximal torus as described above of the gauge group and $\rho>0$ are the positive roots.

Let us briefly describe the analysis of this model for $r \gg 0$.
First, the D terms imply that the $\{ \phi^i_a, \tilde{\phi}^i_a \}$
are not all zero, hence the $\{ B^{i_1 \cdots i_k},
\tilde{B}^{i_1\cdots i_k} \}$ are not all zero.  The F terms imply
\begin{displaymath}
\tilde{B}^{i_1 \cdots i_k} = f^{i_1 \cdots i_k}(B), \: \: \:
p_{i_1 \cdots i_k} \frac{\partial}{\partial \phi^i_a}
f^{i_1 \cdots i_k}(B) = 0 = 
p_{i_1 \cdots i_k} \frac{\partial}{\partial \tilde{\phi}^i_a} 
\tilde{B}^{i_1 \cdots i_k}.
\end{displaymath}
The first F term equation clearly restricts to the intersection, but we still
need to dispense with the $p$'s.  

Next, note that from the D-terms, since the $\{\phi^i_a\}$ and 
$\{\tilde{\phi}^i_{a'}\}$ vevs
each form a set of $k$ linearly independent vectors in ${\mathbb C}^N$,
the dual quantities
\begin{displaymath}
\epsilon^{a_1 \cdots a_k} \phi^{i_2}_{a_2} \cdots \phi^{i_k}_{a_k}, \: \: \:
\epsilon^{a'_1 \cdots a'_k} \tilde{\phi}^{i_2}_{a'_2} \cdots
\tilde{\phi}^{i_k}_{a'_k}
\end{displaymath}
also form two sets of $k$ linearly independent vectors (in a larger
vector space), 
hence the derivatives
\begin{displaymath}
\frac{\partial}{\partial \phi^i_a} B^{i_1 \cdots i_k}, \: \: \:
\frac{\partial}{\partial \tilde{\phi}^i_{a'} }
\tilde{B}^{i_1 \cdots i_k}
\end{displaymath}
also form two sets of linearly independent vectors.
For generic rotations $f^{i_1 \cdots i_k}$,
\begin{displaymath}
\frac{\partial}{\partial \phi^i_a} f^{i_1 \cdots i_k}(B)
= f^{i_1 \cdots i_k}_{j_1 \cdots j_k} 
\frac{\partial}{\partial \phi^i_a} B^{j_1 \cdots j_k}
\end{displaymath}
form a set of $N \choose k$ linearly independent vectors, and so the
F term equations force all the $p_{i_1 \cdots i_k}$ to vanish.
However, for nongeneric choices of $f$, not all the resulting vectors
will be linearly independent, and so one can get noncompact branches of
unconstrained $p$'s, indicating a physical singularity.

To make this more concrete, let us consider the special case that $k=2$.
In this case, $\partial B / \partial \phi \propto \phi^i_a$.
From the D-terms, the $\phi^i_a$ define two linearly independent
vectors in ${\mathbb C}^N$, and the $\tilde{\phi}^i_{a'}$ denote
another two linearly independent vectors in ${\mathbb C}^N$.
For generic rotations, the F-term constraints include
\begin{displaymath}
p_{i_1 i_2} \epsilon^{ab} f^{i_1 i_2}_{j_1 j_2} \phi^{j_2}_b = 0 =
 p_{i_1 i_2} \epsilon^{a' b'}
\tilde{\phi}^{i_2}_{b'}.
\end{displaymath}
Generic rotation matrices $f^{i_1 i_2}_{j_1 j_2}$ can rotate the
$\phi^i_a$ into $N \choose 2$ linearly independent vectors, and in such
cases the only way to satisfy the F-term constraints is for all 
$p_{i_1 i_2} = 0$.

However, for nongeneric rotations $f$, we can have noncompact branches.
Consider for example the case of trivial rotations (again for $k=2$), 
in which $f$ is the
identity.  Identify $\phi^i_a = \tilde{\phi}^i_{a'}$, then omitting the
now-irrelevant $\epsilon^{ab}$, the F-term constraints become
\begin{displaymath}
p_{i_1 i_2} \phi^{i_2}_a = 0.
\end{displaymath}
Suppose for example we are at a point where $\phi^i_a \propto \delta^i_a$,
which is consistent with D-terms.  Then the $p_{ij}$ for $i, j > 2$ are
completely unconstrained, and their vevs can run along a noncompact
branch.  This is consistent with the mathematics result mentioned earlier
that for the trivial rotation, the self-intersection of $G(k,N)$ is singular
as an element of the family.

Now that we have described general aspects of these GLSMs, let us look
at the result for a particular model.
For example, consider $G(2,5)$.  Here,
\begin{eqnarray*}
2 n k & = & 20, \\
\left( \begin{array}{c} n \\ k \end{array} \right) & = & 10,
\end{eqnarray*}
hence for generic maps $f^{i_1 \cdots i_k}$, all the $p$'s will be
forced to vanish, and this GLSM will correctly describe the
corresponding geometry.  However, for nongeneric $f^{i_1 \cdots i_k}$,
this description will break down.  For example, if $f^{i_1 \cdots i_k}$
is the identity, then instead of $2 n k$ independent equations, there are
only $n k = 10$ independent equations, at which point 
linear algebra suggests the $p_{i_1 \cdots i_k}$ may become nonzero, and
our description of the geometry breaks down.

This description of the geometry meshes with the mathematics of
the intersection of $G(2,5)$ with itself.

\subsubsection{Intersection with Veronese embedding}

So far we have discussed the intersection of a single Grassmannian
$G(k,N)$ with its own Pl\"ucker image after a linear deformation.
In principle, we can apply the same ideas to intersections of
different Grassmannians.  Their Pl\"ucker embeddings will land in
projective spaces of different sizes, so one must specify a map between
projective spaces, but given such a map, the rest is straightforward.

Specifically, if we compose the Pl\"ucker embedding of a Grassmannian
$G(k,N)$ with a degree $d$ Veronese map, the result lies in the same
projective space as the Pl\"ucker embedding of the Grassmannian
$G(k',N')$ for
\begin{displaymath}
k' = d, \: \: \:
N' = \left( \begin{array}{c} N \\ k \end{array} \right) - 1 + d.
\end{displaymath}
The intersection can be described
by the GLSM with gauge group
\begin{displaymath}
\frac{
U(1) \times SU(k) \times SU(k')
}{
{\mathbb Z}_k 
},
\end{displaymath}
with matter
\begin{itemize}
\item $N$ chiral superfields in the $({\bf k},{\bf 1})_1$ representation,
\item $N'$ chiral superfields in the $({\bf 1},{\bf k'})_k$ representation,
\item $\left( \begin{array}{c} N' \\ k' \end{array} \right)$ chiral
superfields $p_{i'_1 \cdots i'_{k'}}$ (antisymmetric in their
indices) in the $({\bf 1},{\bf 1})_{- k k'}$
representation,
\end{itemize}
and superpotential
\begin{displaymath}
W \: = \: p_{i'_1 \cdots i'_{k'} } \left(
\tilde{B}^{i'_1 \cdots i'_{k'} } - 
\nu^{i'_1 \cdots i'_{k'}}(B) \right),
\end{displaymath}
with baryons $B^{i_1 \cdots i_k}$, $\tilde{B}^{i'_1 \cdots i'_{k'}}$
constructed from $SU(k)$- and $SU(k')$-invariant combinations of
fundamental-charged chiral superfields, as usual.

In passing, the reader should note that the finite group quotients
are such that the Grassmannian input to the Veronese embedding is
built with a $U(k)$ gauge group, but the Grassmannian intersected
at the output is constructed with a $U(1) \times SU(k')$ gauge
group.  This is closely related to the concern with nonminimally-charged
chiral superfields corresponding to homogeneous coordinates in the
target of the Veronese embedding.  First, since the Lie algebras are the
same in both cases, the perturbative physics is identical; these differences
in finite group quotients can only affect nonperturbative sectors.
Having gauge group $U(1) \times SU(k')$ reflects the fact that the only
allowed maps into the Grassmannian have degree divisible by $k'$, the
degree of the Veronese embedding, as discussed in other
circumstances in {\it e.g.} 
\cite{Pantev:2005zs,Pantev:2005rh,Hellerman:2006zs,Sharpe:2014tca}.

Judging from the $U(1)$ charges, the Calabi-Yau condition is that
\begin{equation}  \label{eq:int-veronese:cy}
N  + N' k' \: = \: 
\left( \begin{array}{c} N' \\ k' \end{array} \right) k'.
\end{equation}
We can see this mathematically as follows.
Let $G = G(k,N)$, $G' = G(k',N')$, and
${\mathbb P}'$ denotes the projective space of dimension
\begin{displaymath}
\left( \begin{array}{c} N' \\ k' \end{array} \right) - 1 .
\end{displaymath}
From the adjunction formula,
\begin{displaymath}
K_{G'} = K_{ {\mathbb P}'} |_{G'}  +
c_1\left( N_{G'/{\mathbb P}'} \right),
\end{displaymath}
and if the intersection $X = G \cap G'$ is transversal,
then we have $N_{X/G} \cong \nu^* N_{G'/{\mathbb P}'}|_X$, hence
\begin{displaymath}
c_1\left( N_{X/G} \right) = d c_1\left( N_{G'/{\mathbb P}'} |_X \right),
\end{displaymath}
where $\nu$ denotes the degree $d$ Veronese embedding.
Using the adjunction formula again, we have
\begin{eqnarray*}
K_X & = & K_G|_X  + c_1\left( N_{X/G} \right), \\
& = & K_G|_X  + d \left[
K_{G'}|_X  -
K_{ {\mathbb P}' }|_X  \right], \\
& = &
N + d \left[ N' - \left( \begin{array}{c} N' \\ k' \end{array} \right) \right],
\end{eqnarray*}
and so we recover the Calabi-Yau condition~(\ref{eq:int-veronese:cy}).

The expected dimension of this intersection is
\begin{equation}
k' (N'-k') + k(N-k) - \left( \begin{array}{c} N' \\ k' \end{array} \right)
+ 1.
\end{equation}
This can be derived from the fact that the expected dimension of
$X \cap Y \subset S$ is
\begin{displaymath}
\dim X + \dim Y - \dim S.
\end{displaymath}

As a result, nontrivial examples are limited in number.  One 
class of examples is the case $d=1$, which describe the intersection
of the Grassmannian $G(k,N)$ with the projective space that defines the
image of its Pl\"ucker embedding.  In other words, the case $d=1$
corresponds to a complicated description of the Grassmannian $G(k,N)$.
We can see this directly in physics as follows.  From our general
prescription, the GLSM is a 
\begin{displaymath}
\frac{ U(1) \times SU(k) \times SU(1) }{ {\mathbb Z}_k } = U(k)
\end{displaymath}
gauge theory with matter
\begin{itemize}
\item $N$ chiral superfields $\phi^i_a$ in the fundamental representation,
\item $N'$ = $N \choose k$ chiral superfields $\tilde{\phi}^{i'} =
\tilde{\phi}^{i_1 \cdots i_k}$ in the
${\bf 1}_k$ representation,
\item $N'$ chiral superfields $p_{i'} = p_{i_1 \cdots i_k}$ in the ${\bf 1}_{-k}$ representation,
\end{itemize}
and superpotential
\begin{displaymath}
W \: = \: p_{i_1 \cdots i_k} \left( \tilde{\phi}^{i_1 \cdots i_k} - 
B^{i_1 \cdots i_k} \right).
\end{displaymath}
Integrating out the $p$'s and $\tilde{\phi}$'s immediately recovers
the Grassmannian $G(k,N)$.

Another class of examples has $d=2$ and $k=1$.
In these examples, one intersects the Pl\"ucker embedding of the
Grassmannian $G(2,N+1)$ with the Veronese image of $G(1,N) = 
{\mathbb P}^{N-1}$.  For $N=2, 3$, these have expected dimension $1$.

\subsubsection{Intersection with Segre embedding}

For another example, suppose we have three Grassmannians $G(k_1,N_1)$,
$G(k_2,N_2)$, $G(k_3,N_3)$ where the $k$'s and $N$'s obey
\begin{displaymath}
\left( \begin{array}{c} N_3 \\ k_3 \end{array} \right) \: = \:
\left( \begin{array}{c} N_1 \\ k_1 \end{array} \right)
\left( \begin{array}{c} N_2 \\ k_2 \end{array} \right).
\end{displaymath}
Consider the intersection of the Pl\"ucker embedding of $G(k_3,N_3)$ with the
Segre embedding of the images of the Pl\"ucker embeddings of 
$G(k_1,N_1)$ and $G(k_2,N_2)$.  We can realize this in a GLSM as follows.
Consider the GLSM with gauge group
\begin{displaymath}
U(k_1) \times U(k_2) \times SU(k_3),
\end{displaymath}
with matter
\begin{itemize}
\item $N_1$ chiral multiplets in representation $({\bf k_1},{\bf 1},{\bf 1})$,
\item $N_2$ chiral multiplets in representation $({\bf 1},{\bf k_2},{\bf 1})$,
\item $N_3$ chiral multiplets in the fundamental of $SU(k_3)$, of charge $k_1$
under $\det U(k_1)$ and charge $k_2$ under $\det U(k_2)$,
\item $\left( \begin{array}{c} N_3 \\ k_3 \end{array} \right)$
chiral multiplets $p_{i_1 \cdots i_{k3}}$ 
of charge $-k_1$ under $\det U(k_1)$ and charge $-k_2$ under $\det U(k_2)$,
\end{itemize}
and superpotential
\begin{displaymath}
W \: = \: p_{i_1 \cdots i_{k3}} \left( B_3^{i_1 \cdots i_{k3}} - 
S^{i_1 \cdots i_{k3}}(B_1, B_2) \right),
\end{displaymath}
where $B_i$ denotes baryons for $SU(k_i)$ ({\it i.e.} the Pl\"ucker images
of each Grassmannian), and $S^{i_1 \cdots i_{k3}}$ denotes the image of
the Segre embedding map.

In passing, the fact that there is no ${\mathbb Z}_{k_3}$ finite
group quotient in the gauge group reflects the fact that maps into the
third Grassmannian are of degree satisfying certain divisibility conditions,
as we have seen previously in discussions of GLSM realizations of
Veronese embeddings.

It is straightforward to find (not necessarily Calabi-Yau) examples,
from identities such as
\begin{displaymath}
\left( \begin{array}{c} 6 \\ 1 \end{array} \right)
\left( \begin{array}{c} 6 \\ 3 \end{array} \right)
\: = \:
\left( \begin{array}{c} 10 \\ 3 \end{array} \right).
\end{displaymath}

It is straightforward to check from $U(1)$ charges that the
Calabi-Yau condition is 
\begin{displaymath}
N_1 +N_3 =  \left( \begin{array}{c} N_3 \\ k_3 \end{array} \right),
\: \: \:
N_2 +N_3 =  \left( \begin{array}{c} N_3 \\ k_3 \end{array} \right).
\end{displaymath}
We can also derive this mathematically as follows.
Let $X = G(k_1,N_1) \times G(k_2,N_2)$, $Y = G(k_3,N_3)$,
$S$ the projective space in which $Y$ is embedded via Pl\"ucker,
and $W$ the intersection.
The Calabi-Yau condition can then be stated as
\begin{displaymath}
K_X |_W + K_Y|_X = K_S|_W.
\end{displaymath}
In the present case,
\begin{displaymath}
K_X = (-N_1,-N_2), \: \: \: K_Y = -N_3.
\end{displaymath}
The Calabi-Yau condition then becomes the statement derived above from
$U(1)$ charges.

\subsection{Intersections of Pfaffians}
\label{sect:int-pfaff}

In this section, we will describe perturbative GLSMs for intersections of
Pfaffians, following the PAX and PAXY
models described in \cite{Jockers:2012zr}.

\subsubsection{PAX models}

First, let us recall the PAX model of a GLSM for a Pfaffian.
Suppose the Pfaffian variety we wish to describe arises as the locus
$\{ {\rm rank}\, A(\phi) \leq k\} \subset {\mathbb P}^n$, 
for $A(\phi)$ an $N \times N$ matrix whose entries
are homogeneous in the homogeneous coordinates of a projective 
space ${\mathbb P}^n$.  For simplicity, let us assume that each entry
in the matrix $A$ has degree $d$.

In the PAX model, a GLSM describing this variety is given as a 
$U(1) \times U(N-k)$ gauge
theory with the following matter:
\begin{itemize}
\item $n+1$ chiral superfields $\phi_i$, of charge $1$ under
the $U(1)$ and neutral under $U(N-k)$ (corresponding to homogeneous
coordinates on ${\mathbb P}^n$), 
\item $N$ chiral superfields $x_{i a}$
transforming in the ${\bf \overline{N-k}}$ representation of $U(N-k)$
(and neutral under the $U(1)$), 
\item and $N$ chiral superfields
$p_i^a$, transforming in the ${\bf N-k}$ representation of $U(N-k)$
and of charge $-d$ under the $U(1)$, 
\end{itemize}
with superpotential
\begin{displaymath}
W \: = \: \sum_{i=1}^N \sum_{a=1}^{N-k} p_i^a A^{ij}(\phi) x_{j a}.
\end{displaymath}
(The name of the model follows from the form of the superpotential.)

We could describe the intersection of two loci as follows.
Suppose we have an $N_1 \times N_1$ matrix $A^{ij}(\phi)$, entries
homogeneous of degree $d_1$ in the homogeneous coordinates on 
${\mathbb P}^n$, and an $N_2 \times N_2$ matrix $B^{ij}(\phi)$,
entries homogeneous of degree $d_2$ in the homogeneous coordinates on
the same ${\mathbb P}^n$, and we want to describe the intersection
\begin{displaymath}
\{ {\rm rank}\, A \leq k_1 \} \cap 
\{ {\rm rank}\, B \leq k_2 \} \: \subset \: {\mathbb P}^n.
\end{displaymath}
To do this, we will use two copies of the PAX model.  Specifically,
consider the GLSM defined by a $U(1) \times U(N_1 - k_1) \times
U(N_2 - k_2)$ gauge theory with the following matter:
\begin{itemize}
\item $n+1$ chiral superfields $\phi_i$ of charge $1$ under the $U(1)$
and neutral under $U(N_1-k_1) \times U(N_2 - k_2)$ (corresponding to
homogeneous coordinates on ${\mathbb P}^n$),
\item $n+1$ chiral superfields $\tilde{\phi}_i$ of charge $1$ under
the $U(1)$ and neutral under $U(N_1-k_1) \times U(N_2-k_2)$,
corresponding to homogeneous coordinates on a second copy of
${\mathbb P}^n$,
\item $n+1$ chiral superfields $q_i$ of charge $-1$ under
the $U(1)$ and neutral under $U(N_1-k_1) \times U(N_2-k_2)$,
acting as Lagrange multipliers identifying the two sets of homogeneous
coordinates,
\item $N_1$ chiral superfields $x_{i a}$ transforming in the
${\bf ( \overline{N_1 - k_1}, 1) }$ representation of
$U(N_1 - k_1) \times U(N_2 - k_2)$ (and neutral under the $U(1)$),
\item $N_1$ chiral superfields $p_i^a$, transforming in the
${\bf (N_1-k_1),1)}$ representation of $U(N_1-k_1) \times U(N_2-k_2)$
and of charge $-d_1$ under the $U(1)$,
\item $N_2$ chiral superfields $\tilde{x}_{i' a'}$ transforming in the
${\bf (1, \overline{N_2-k_2}) }$ representation of
$U(N_1-k_1) \times U(N_2-k_2)$ (and neutral under the $U(1)$),
\item $N_2$ chiral superfields $\tilde{p}_{i'}^{a'}$,
transforming in the ${\bf (1,N_2-k_2)}$ representation of
$U(N_1-k_1) \times U(N_2-k_2)$ and of charge $-d_2$ under the $U(1)$,
\end{itemize}
with superpotential
\begin{displaymath}
W \: = \: \sum_{i=1}^{N_1} \sum_{a=1}^{N_1-k_1} p_i^a A^{ij}(\phi)
x_{j a} \: + \:
\sum_{i'=1}^{N_2} \sum_{a'=1}^{N_2-k_2} \tilde{p}_{i'}^{a'}
B^{i' j'}(\tilde{\phi}) \tilde{x}_{j' a'} \: + \:
\sum_{i=1}^{n+1} q_i( \phi_i - \tilde{\phi}_i).
\end{displaymath}

Next, for simplicity, we will assume that the $q_i$ have been
integrated out, so that we can identify $\phi_i = \tilde{\phi}_i$.

Now, let us carefully analyze this theory, closely following 
\cite{Jockers:2012zr}[section 3.2].  We shall focus on D- and F-terms
for the extra data above the base toric variety, so for example we
assume FI parameters are such that not all the $\phi$ vanish.
D-terms for $U(N_1-k_1)$, $U(N_2-k_2)$ imply
\begin{displaymath}
p_i^a (p^{\dag})^i_b - (x^{\dag})^{i a} x_{i b} = r_1 \delta^a_b, \: \: \:
\tilde{p}_i^{a'} (\tilde{p}^{\dag})^i_{b'} - 
(\tilde{x}^{\dag})^{i a'} \tilde{x}_{i b'} = r_2 \delta^{a'}_{b'}.
\end{displaymath}
Define $E^i_a(\phi,x) = A(\phi)^{ij} x_{ja}$,
$\tilde{E}^{i'}_{a'} = B(\phi)^{i'j'} \tilde{x}_{j' a'}$.
The F-term conditions can then be expressed as
\begin{displaymath}
A(\phi)^{ij} x_{ja} = 0, \: \: \:
\tilde{E}^{i'}_{a'} = B(\phi)^{i'j'} \tilde{x}_{j' a'} = 0,
\end{displaymath}
\begin{displaymath}
p_i^a \frac{\partial E^i_a }{\partial (\phi, x, \tilde{x}) } +
\tilde{p}_{i'}^{a'} \frac{\partial \tilde{E}^{i'}_{a'} }{\partial
(\phi, x, \tilde{x}) } = 0.
\end{displaymath}
The first two conditions demonstrate that the vacua lie along the
intersection
\begin{displaymath}
\{ (\phi, x)\, | \, A(\phi) x = 0 \} \cap
\{ (\phi, \tilde{x}) \, | \, B(\phi) \tilde{x} = 0 \}.
\end{displaymath}
Each factor is a (resolution of) a Pfaffian describing the locus where
the rank of a matrix ($A$ or $B$) is bounded by $k$.
Following standard methods, the derivatives $\partial E$, 
$\partial \tilde{E}$ each span the normal bundle to each factor.
The set of such is linearly independent if the intersection is smooth,
and so we see from the last set of F-term conditions
that smoothness implies that $p = \tilde{p} = 0$.

\subsubsection{PAXY models}   \label{sect:int-pf:paxy}

Let us now consider analogous constructions involving the PAXY models
\cite{Jockers:2012zr},
which will play an important role in understanding dual theories later
in this paper.

The PAXY model that describes the locus on which
an $n \times n$ matrix $A(\phi)$ can be factored into a product $Y X$ of two
matrices (one $n \times k$, the other $k \times n$) can be described by a 
GLSM with gauge group $U(k)$ and matter as follows:
\begin{itemize}
\item chiral superfields $\Phi$, charged under a different gauge factor,
describing the space in which the Pfaffian lives,
\item $n$ chiral superfields in the fundamental of $U(k)$,
defining a $k \times n$ matrix $X$,
\item $n$ chiral superfields in the antifundamental of $U(k)$,
defining a $n \times k$ matrix $Y$,
\item $n^2$ chiral superfields forming an $n \times n$ matrix $P$,
\end{itemize}
and superpotential
\begin{displaymath}
W \: = \: \sum_{i,j=1}^n P_{ji} \left( A(\Phi)_{ij} \: - \:
\sum_{a = 1}^k Y_{ia} X_{aj} \right).
\end{displaymath}

To describe the intersection of two closely analogous Pfaffians,
defined by matrices $A(\Phi)$, $B(\Phi)$, each with the same factorization
constraint, we could use a $U(k) \times U(k)$ gauge theory with matter
\begin{itemize}
\item chiral superfields $\Phi$, charged under a different gauge factor,
describing the space in which the intersection of Pfaffians lives,
\item two sets of $n$ chiral superfields, each set in the fundamental of 
a $U(k)$ factor,
defining $k \times n$ matrices $X$, $\tilde{X}$
\item two sets of $n$ chiral superfields, each in the antifundamental of 
a $U(k)$ factor,
defining $n \times k$ matrices $Y$, $\tilde{Y}$,
\item two sets of $n^2$ chiral superfields forming two $n \times n$ matrices 
$P$, $\tilde{P}$,
\end{itemize}
and superpotential
\begin{displaymath}
W \: = \: \sum_{i,j=1}^n P_{ji} \left( A(\Phi)_{ij} \: - \:
\sum_{a = 1}^k Y_{ia} X_{aj} \right) \: + \:
\sum_{i,j=1}^n \tilde{P}_{ji} \left( B(\Phi)_{ij} \: - \:
\sum_{a' = 1}^k \tilde{Y}_{ia'} \tilde{X}_{a'j} \right).
\end{displaymath}

Let us work systematically through the analysis.  (We will closely
follow \cite{Jockers:2012zr}[section 3.3].)  We assume that there
is a phase in which D terms imply that the $\Phi$'s are not all zero, and work
in that phase.  The D-term constraint for the $U(k)$ is
\begin{displaymath}
X_{a i} X^{\dag}_{i b} - Y^{\dag}_{a i} Y_{i b} = r \delta_{ab}, \: \: \:
\tilde{X}_{a' i} \tilde{X}^{\dag}_{i b'} - \tilde{Y}^{\dag}_{a' i} 
\tilde{Y}_{i b'}
= \tilde{r} \delta_{a' b'}.
\end{displaymath}
Assuming that $r \gg 0$, $\tilde{r} \gg 0$,
this forces $X$ and $\tilde{X}$ to have rank $k$.
The F-term constraints have the form
\begin{displaymath}
A(\Phi)_{ij} = \sum_{a = 1}^k Y_{ia} X_{aj} , \: \: \:
B(\Phi)_{ij} = \sum_{a' = 1}^k \tilde{Y}_{ia'} \tilde{X}_{a'j} ,
\end{displaymath}
\begin{displaymath}
P_{ji} Y_{ia} = 0, \: \: \:
P_{ji} X_{aj} = 0, \: \: \:
\tilde{P}_{ji} \tilde{Y}_{ia} = 0, \: \: \:
\tilde{P}_{ji} \tilde{X}_{aj} = 0,
\end{displaymath}
\begin{displaymath}
P_{ji} \frac{\partial A(\Phi)_{ij} }{\partial \Phi_{\alpha} } \: + \:
\tilde{P}_{ji} \frac{\partial B(\Phi)_{ij} }{\partial \Phi_{\alpha} }
\: = \: 0.
\end{displaymath}
With the exception of the last constraint, the F-term constraints above
are identical to those appearing in two copies of the PAXY model,
as analyzed in \cite{Jockers:2012zr}[section 3.3].
It will be helpful to define
\begin{displaymath}
E_{ij} = A(\Phi)_{ij} - \sum_{a = 1}^k Y_{ia} X_{aj} , \: \: \:
\tilde{E}_{ij} = B(\Phi)_{ij} - \sum_{a' = 1}^k \tilde{Y}_{ia'} \tilde{X}_{a'j},
\end{displaymath}
so that the remaining F-term conditions,
beyond $E = 0 = \tilde{E}$, can all be written
as
\begin{displaymath}
P_{ji} \frac{ \partial E(\Phi,X,Y,\tilde{X},\tilde{Y})_{ij} }{
\partial (\Phi, X, Y, \tilde{X}, \tilde{Y}) } +
\tilde{P}_{ji} \frac{\partial \tilde{E}(\Phi, \tilde{X}, \tilde{Y})_{ij}
}{\partial (\Phi, X, Y, \tilde{X}, \tilde{Y}) } = 0.
\end{displaymath}
As in the discussion of the PAX model,
the derivatives $\partial E$, $\partial \tilde{E}$ span the normal
bundle.  These will be linearly independent if the intersection is smooth,
so we see from the last set of F-term conditions
that smoothness implies that $P = 0 = \tilde{P}$.

\subsubsection{PAXY for antisymmetric matrices}

Later in this paper we'll see examples of intersections of Pfaffians
for antisymmetric matrices.  Suppose we have $N \times N$
antisymmetric
matrices $A$, $B$, each with entries linear in the homogeneous coordinates
of ${\mathbb P}^n$, and we want to compute the intersection of Pfaffians
defined by the rank $k$ locus of each matrix.  We assume $N$ is odd,
and then of necessity $k$ is even.

The GLSM describing the Pfaffian for a single antisymmetric matrix $A$,
of the form above, following \cite{Jockers:2012zr}[section 3.5],
is defined by a
\begin{displaymath}
\frac{
U(1) \times USp(k)
}{
{\mathbb Z}_2
}
\end{displaymath}
gauge theory with matter
\begin{itemize}
\item $n+1$ chiral superfields $\phi$, forming the homogeneous
coordinates on the projective space, of charge $2$ under the $U(1)$
and neutral under $USp(k)$,
\item $N$ chiral superfields $x_{ia}$ in the fundamental of $USp(k)$,
and of charge $1$ under the $U(1)$,
\item $N^2$ chiral superfields $p_{ij}$ of charge $-2$ under the $U(1)$
and neutral under $USp(k)$,
\end{itemize}
with superpotential
\begin{displaymath}
W \: = \: \sum_{ij} p_{ij} \left( A(\phi)_{ij} - x_{ia} x_{jb} J^{ab}
\right),
\end{displaymath}
where $J$ is the $k \times k$ antisymmetric matrix in block form
\begin{displaymath}
J \: = \: - \left[ \begin{array}{cc}
0 & 1 \\ -1 & 0 \end{array} \right].
\end{displaymath}

With that in mind, it should be clear that the intersection of two
such Pfaffians, defined by antisymmetric $N \times N$ matrices $A$, $B$,
each linear in the homogeneous coordinates of ${\mathbb P}^n$, is
defined by a
\begin{displaymath}
\frac{
U(1) \times USp(k) \times USp(k)
}{
{\mathbb Z}_2 \times {\mathbb Z}_2
}
\end{displaymath}
gauge theory with matter
\begin{itemize}
\item $n+1$ chiral superfields $\phi$, forming the homogeneous
coordinates on the projective space, in representation $({\bf 1},{\bf 1})_2$
of the gauge group,
\item $N$ chiral superfields $x_{ia}$ in the representation
$({\bf k},{\bf 1})_1$ of the gauge group,
\item $N$ chiral superfields $\tilde{x}_{i' a'}$ in the representation
$({\bf 1},{\bf k})_1$ of the gauge group,
\item $N^2$ chiral superfields $p_{ij}$ in the representation
$({\bf 1},{\bf 1})_{-2}$,
\item $N^2$ chiral superfields $\tilde{p}_{i' j'}$ in the representation
$({\bf 1},{\bf 1})_{-2}$,
\end{itemize}
with superpotential
\begin{displaymath}
W \: = \: \sum_{ij} p_{ij} \left( A(\phi)_{ij} - x_{ia} x_{jb} J^{ab} \right)
\: + \: 
\sum_{i' j'} \tilde{p}_{i' j'} \left( B(\phi)_{i' j'} - 
\tilde{x}_{i' a'} \tilde{x}_{j' b'} J^{a' b'} \right).
\end{displaymath}

It is straightforward to check that the sum of the $U(1)$ charges will
vanish when
\begin{displaymath}
n+1 \: = \: N(N-k).
\end{displaymath}

We can derive this Calabi-Yau condition mathematically as follows.
First, consider an $N \times N$
skew-symmetric matrix, $N$ odd, of generic variables.
There are
\begin{displaymath}
M = \left( \begin{array}{c} N \\ 2 \end{array} \right)
\end{displaymath}
such variables.  Let $X$ denote the locus in ${\mathbb P} \equiv
{\mathbb P}^{M-1}$ where
the rank of this matrix is less than $k$.  The structure sheaf
${\cal O}_X$ admits a resolution of the form
\cite{weyman}[theorem 6.4.1(c)]
\begin{displaymath}
0 \longrightarrow {\cal O}_{\mathbb P}(-N(N-k)/2) \longrightarrow \cdots
\longrightarrow {\cal O}_{\mathbb P}
\end{displaymath}
If we specialize the generic variables to linear forms in
${\mathbb P}^n$, then the resolution remains exact if and only if the
codimension is preserved.  In that case, the resolution above stays the
same, so by adjunction, the canonical bundle of the Pfaffian variety
$X$ is ${\cal O}_X(N(N-k)/2 - n - 1)$.

Now, for the intersection of two such Pfaffians, call them $X$ and $Y$,
to be Calabi-Yau, one must require
\begin{displaymath}
K_X + K_Y = K_{\mathbb P},
\end{displaymath}
or explicitly
\begin{displaymath}
2( N (N-k) - n - 1) = - (n+1),
\end{displaymath}
hence $N(N-k)=n+1$, the condition derived above from $U(1)$ charges.

\section{Phases of GLSMs for self-intersection of $G(2,N)$ for $N$ odd}
\label{sect:phases}

In this section we will study the GLSM for the self-intersection of
two copies of $G(2,N)$, one linearly rotated.  We previously discussed
this setup in section~\ref{sec:int-grass-rot}.
We will see that the $r \ll 0$ phase
of this GLSM describes an analogous intersection of Pfaffians, and will
verify this interpretation by dualizing the GLSM and examining the phases
of the dual.  

First, let us recall the condition~(\ref{eq:grass-nonempty})
for such self-intersections to
be nonempty.  In the present case of $G(2,N)$ for $N$ odd, it is
straightforward to check that the only nonempty cases are those for
which $N \leq 7$.  Furthermore, from the expected dimension
formula~(\ref{eq:int-grass:dim}), the case $N=3$ has dimension 2,
the case $N=5$ has dimension 3, and the case $N=7$ has dimension 0.

Of these three cases, the case in which we are primarily interested
is the case $N=5$.  This case corresponds to a Calabi-Yau threefold:
from equation~(\ref{eq:int-grass:cy}), we see that this intersection is
be Calabi-Yau, and
for generic deformations, the formula for the expected
dimension~(\ref{eq:int-grass:dim}) tells us that this is a threefold.

This example was studied mathematically in \cite{bcp,or}.
They argued that this Calabi-Yau threefold,
also studied in \cite{kan}, with Hodge numbers $h^{1,1}=1$ and
$h^{2,1}=51$, is deformation-equivalent and derived
equivalent, but not birational, to an intersection of dual Grassmannians,
which are equivalent to an intersection of two Pfaffians.

We shall see
that this derived equivalence is reflected in the structure of the phases
of our GLSM:  the $r \gg 0$ phase of our GLSM will describe the 
self-intersection of $G(2,5)$, and the $r \ll 0$ phase will describe the
self-intersection of the dual Grassmannians, realized as Pfaffian
varieties. 
The fact that they both appear as phases of the same GLSM provides a
physics argument for their derived equivalence, as the B model is
independent of K\"ahler moduli. 

In fact, the structure of these phases will be of further interest:
in the first GLSM we study, the geometry of the $r \gg 0$ phase is
realized perturbatively, but the geometry of the $r \ll 0$ phase
is realized nonperturbatively via the insights of \cite{Hori:2006dk}.
In the next subsection, we dualize both $SU(2)$ gauge factors,
and find the opposite:  the $r \gg 0$ phase realizes the same geometry
(self-intersection of $G(2,5)$), but does so nonperturbatively, whereas
the $r \ll 0$ phase realizes the intersection of two Pfaffians perturbatively.
Finally, in the subsequent subsection, we dualize on a single nonabelian
gauge factor, and find that both phases realize geometry through a mix
of perturbative effects (in one gauge factor) and nonperturbative effects
(in the other gauge factor).

Although the $N=5$ case will be our primary interest, to be thorough
we will outline the analysis for the cases $N=3, 7$ as well.

In passing, since the gauge group 
can also be written in the form $U(k) \times U(k)$,
each with $N$ fundamentals, we could also dualize each factor to
$U(N-k)$, following the duality prototyped by $G(k,N) \cong G(N-k,N)$.
In the present case, for $k=2,N=5$ for example,
invoking this duality on either gauge factor separately would
yield the intersection $G(2,5) \cap G(3,5)$ and phases thereof;
dualizing on both factors simultaneously would yield the intersection
$G(3,5) \cap G(3,5)$ and phases thereof.  In either case, the resulting
theories would be functionally identical to those considered here.
We thank S.~Galkin for pointing this out to us.

\subsection{First GLSM}

The gauge group for this GLSM is 
\begin{displaymath}
\frac{ U(1) \times SU(2) \times SU(2) }{ {\mathbb Z}_2 \times {\mathbb Z}_2 },
\end{displaymath}
where each ${\mathbb Z}_2$ relates the center of one $SU(2)$ to a subgroup
of $U(1)$.  The matter consists of
\begin{itemize}
\item $N$ chiral multiplets $\phi^i_a$ in the ${\bf (2,1)_1}$ representation
of the gauge group,
\item $N$ chiral multiplets $\tilde{\phi}^j_{a'}$ in the ${\bf (1,2)_1}$
representation of the gauge group, and
\item $(1/2) N (N-1)$ chiral multiplets $p_{i_1 i_2} = - p_{i_2 i_1}$
in the ${\bf (1,1)_{-2}}$ representation of the gauge group,
\end{itemize}
with superpotential
\begin{displaymath}
W \: = \: \sum_{i_1 < i_2} p_{i_1 i_2} \left(
f^{i_1 i_2}(B) - \tilde{B}^{i_1 i_2} \right),
\end{displaymath}
where
\begin{displaymath}
B^{i_1 i_2} = \epsilon^{ab} \phi^{i_1}_{a} \phi^{i_2}_b, \: \: \:
\tilde{B}^{i_1 i_2} = \epsilon^{a' b'} \tilde{\phi}^{i_1}_{a'}
\tilde{\phi}^{i_2}_{b'},
\end{displaymath}
are the `baryons' in each of the two $SU(2)$ factors of the gauge group
(which mathematically define the homogeneous coordinates of the 
Pl\"ucker embedding $G(2,N) \hookrightarrow {\mathbb P}^{(1/2) N (N-1) - 1}$),
and $f^{i_1 i_2}(x) = - f^{i_2 i_1}(x)$ are a set of ten linear
homogeneous polynomials in the homogeneous coordinates of 
${\mathbb P}^9$.  In effect, $B^{i_1 i_2}$ and $\tilde{B}^{i_1 i_2}$
define the Pl\"ucker embedding $G(2,N) \hookrightarrow {\mathbb P}^{(1/2) N (N-1) - 1}$, and
$f^{i_1 i_2}$ describes the deformation of one of those embeddings.

For $r \gg 0$, the analysis of this GLSM is straightforward, and the result
is the intersection of two copies of $G(2,N) \hookrightarrow {\mathbb P}^{
(1/2) N (N-1) - 1}$, one deformed by the linear maps $f^{i_1 i_2}$, as enforced
by the Lagrange multipliers $p_{i_1 i_2}$.

Now, let us turn to the $r \ll 0$ phase.  We write the superpotential
in the form
\begin{displaymath}
W \: = \: A(p)_{i_1 i_2} \phi^{i_1}_a \phi^{i_2}_b \epsilon^{ab} \: + \:
C(p)_{i_1 i_2} \tilde{\phi}^{i_1}_{a'} \tilde{\phi}^{i_2}_{b'} \epsilon^{a' b'},
\end{displaymath}
where $C(p)_{i_1 i_2} = p_{i_1 i_2}$.  In this phase, the D terms prevent
the $p_{i_1 i_2}$ from all vanishing simultaneously, and so they form homogeneous
coordinates on ${\mathbb P}^{ (1/2) N (N-1) - 1}$.  The analysis of 
this phase is then closely related to that described for the
R{\o}dland model in \cite{Hori:2006dk}.  As in the analysis there,
we work in a Born-Oppenheimer 
approximation on the space of $p$'s.  Since
$A(p)$ is an antisymmetric $N \times N$ matrix, where $N$ is odd,
and antisymmetric matrices have even rank, its rank $n$ must be one of
$N-1, N-3, \cdots, 0$, and since $A(p)$ acts as a mass matrix,
on loci for which $A(p)$ has rank $n$, there are $N-n$ massless doublets.  
On loci for which it has rank $N-1$, there is one
massless $\phi$ doublet in the first $SU(2)$, and the vacua all run to infinity,
which would leave no vacua in the IR,
following \cite{Hori:2006dk}.  If the rank of $A(p)$ is $N-3$, then there
are three massless doublets of the first $SU(2)$, leaving one vacuum
in the IR \cite{Hori:2006dk}.  Similarly, only loci with $C(p)$ of
rank no greater than $N-3$ have vacua in the second $SU(2)$.

Putting this together, we see that the $r \ll 0$ limit should be interpreted
as the intersection of two Pfaffians in ${\mathbb P}^{ (1/2) N (N-1) - 1}$,
where $A(p)$ has rank $\leq N-3$ and $C(p)$ has rank $\leq N-3$.

The case $N=5$ is particularly interesting.  In this case, the two
Pfaffians intersecting above are both equivalent\footnote{
See for example \cite{Donagi:2007hi}[section 4.2.2] 
for a discussion for physicists of this special case of Pfaffians
and their relation to Grassmannians.
} to Grassmannians
$G(2,5)$, or more precisely, dual Grassmannians to those appearing in the
$r \gg 0$ phase.  In the notation of \cite{bcp}, the $r \gg 0$ phase
describes $X = G(2,5) \cap G(2,5)$, and the $r \ll 0$ phase describes
$Y$, given as the intersection of the dual Grassmannians.  In \cite{bcp},
$X$ and $Y$ were described as `Kanazawa double mirrors,' so-named because
their mirrors were believed to be related by complex structure deformation.
Here, we see that $X$ and $Y$ arise as different K\"ahler phases of
the same GLSM -- in particular, related by K\"ahler deformations, and so
indeed their mirrors are necessarily related by complex structure deformations.

Using the parametrization of the maximal torus as discussed in 
section~\ref{sec:int-grass-rot} the effective potential on the Coulomb
branch is
\begin{eqnarray*}
  \widetilde{\mathcal{W}}_{eff}&=&-r(\alpha_1+\alpha_2)+\pi
  i(\alpha_1-\alpha_2)+\pi
  i(2\beta_1-\alpha_1-\alpha_2)\nonumber\\
  &&+10(\alpha_1+\alpha_2)(\log(\alpha_1+\alpha_1)-1)
  -5\alpha_1(\log\alpha_1-1)-5\alpha_2(\log\alpha_2-1)\nonumber\\
  &&-5\beta_1(\log\beta_1-1)
  -5(\alpha_1+\alpha_2-\beta_1)(\log(\alpha_1+\alpha_2-\beta_1)-1)
\end{eqnarray*}
Here we have eliminated the coordinate $\beta_2$. Computing the
critical locus of this and removing any solutions that are fixed under
the Weyl group action ({\it i.e.} $\alpha_1=\alpha_2$, $\beta_1=\beta_2$) we
find three singular points:
\begin{displaymath}
  e^{-t}=-1,\; \frac{1}{(-1+\omega)^{10}},\; \frac{1}{(1+\omega^2)^{10}} 
  \qquad \omega=e^{\frac{2\pi i}{10}}.
\end{displaymath}
This is in agreement with the differential operator AESZ 101 in \cite{vanstraten}.

\subsection{Dual description}

We now describe a dual GLSM, created by dualizing both $SU(2)$ factors.

From \cite{Hori:2011pd}[section 5.6], 
for an odd number $N \geq k+3$ massless fundamentals,
there is a duality between the $Sp(k)=USp(k)$ gauge theory with
$N$
fundamentals $\phi_i$
and $Sp(N-k-1)$ gauge theory with $N$ fundamentals $\varphi_i$ and
$(1/2)N(N-1)$ singlets $b_{ij} = - b_{ji}$ (dual to the baryons of the
original theory), with superpotential
\begin{displaymath}
W \: = \: \sum_{ij} b_{ij} [ \varphi_i \varphi_j ] \: = \:
\sum_{ij} \sum_{ab} b_{ij} \varphi_i^a \varphi_j^b J_{ab},
\end{displaymath}
where $J$ is an antisymmetric $(N-k-1)\times(N-k-1)$ matrix of the block form
\begin{displaymath}
J \: = \: \left[ \begin{array}{cc} 0 & 1 \\ -1 & 0 \end{array} \right],
\end{displaymath}
as commonly arises in symplectic geometry.

Here, so long as $N \geq 5$, we can follow this prescription and dualize
the first $SU(2)$ with its $N$ fundamentals $\phi^i_a$ to an
$Sp(N-3)$ gauge theory with $N$ fundamentals $\varphi_i^a$ and
$(1/2) N (N-1)$ singlets $b^{ij} = - b^{ji}$.  The second $SU(2)$ dualizes to
an $Sp(N-3)$ gauge theory with $N$ fundamentals $\tilde{\varphi}_i^{a'}$
and singlets $\tilde{b}^{ij}$.  The $U(1)$ factor is unchanged, and as the
center of $Sp(n)$ is\footnote{
See for example \cite{Distler:2007av}[appendix A]
for a more exhaustive discussion of centers of
Lie groups.
} ${\mathbb Z}_2$, the finite group quotient is also
unchanged, hence the gauge group of the dual theory is
\begin{displaymath}
\frac{ U(1) \times Sp(N-3) \times Sp(N-3) }{ {\mathbb Z}_2  \times
{\mathbb Z}_2 }.
\end{displaymath}

The dual superpotential has the form
\begin{eqnarray*}
W & = & \left( A(p)_{i_1 i_2} + \varphi_{i_1}^a \varphi_{i_2}^b J_{ab}
\right) b^{i_1 i_2} \: + \:
\left( C(p)_{i_1 i_2} + \tilde{\varphi}_{i_1}^{a'} \tilde{\varphi}_{i_2}^{b'}
J_{a' b'} \right) \tilde{b}^{i_1 i_2}, \\
& = &
\sum_{i_1 < i_2} p_{i_1 i_2} \left( f^{i_1 i_2}(b) - \tilde{b}^{i_1 i_2}
\right) \: + \:
\varphi_{i_1}^a \varphi_{i_2}^b J_{ab} b^{i_1 i_2} \: + \:
\tilde{\varphi}_{i_1}^{a'} \tilde{\varphi}_{i_2}^{b'} J_{a' b'}
\tilde{b}^{i_1 i_2}.
\end{eqnarray*}

Since the $b^{ij}$ and $\tilde{b}^{ij}$ are dual to baryons, they have
gauge $U(1)$ charge $2$.  The form of superpotential couplings implies that
$\varphi_i^a$, $\tilde{\varphi}_i^{a'}$ have $U(1)$ charge $-1$.  Finally,
the $p_{ij}$ have the same $U(1)$ charge as in the original theory,
namely $-2$.

From counting $U(1)$ charges, it is straightforward to check that this
intersection will be Calabi-Yau only in the special case that $N=5$,
as expected.

In the phase $r \gg 0$, D terms imply that
the $b^{ij}$ and $\tilde{b}^{ij}$ are not all
zero.
The pertinent form of the superpotential is
\begin{displaymath}
W \: = \:
\sum_{i_1 < i_2} p_{i_1 i_2} \left( f^{i_1 i_2}(b) - \tilde{b}^{i_1 i_2}
\right) \: + \:
\varphi_{i_1}^a \varphi_{i_2}^b J_{ab} b^{i_1 i_2} \: + \:
\tilde{\varphi}_{i_1}^{a'} \tilde{\varphi}_{i_2}^{b'} J_{a' b'}
\tilde{b}^{i_1 i_2}.
\end{displaymath}
In the previous GLSM, the $r \gg 0$ phase could be understood perturbatively,
but in this model, we must analyze this phase nonperturbatively,
following \cite{Hori:2006dk},  
combined with the same assumptions about $f$ as in the original model.  We work in a Born-Oppenheimer approximation
across the space of $b$'s, $\tilde{b}$'s,
and interpret the terms
\begin{displaymath}
\varphi_{i_1}^a \varphi_{i_2}^b J_{ab} b^{i_1 i_2} \: + \:
\tilde{\varphi}_{i_1}^{a'} \tilde{\varphi}_{i_2}^{b'} J_{a' b'}
\tilde{b}^{i_1 i_2}
\end{displaymath}
as $b$- and $\tilde{b}$-dependent mass matrices for $\varphi$ and
$\tilde{\varphi}$. 

For the case $N=5$, the gauge group has two $SU(2)$ factors, and so
we can apply the results of \cite{Hori:2006dk} to argue that the
only contribution to the vacua will arise from the locus where the
(antisymmetric) $N \times N$ mass matrices have rank 2, so that there
are three massless doublets and hence one vacuum in the IR.
(On the rank 4 locus, there is one massless doublet, and no vacua in the
IR \cite{Hori:2006dk}.)  Thus, from each $SU(2)$ factor, we get a 
Pfaffian variety, over the (cone over) a copy of ${\mathbb P}^9$ determined
by $(b^{ij})$ or $(\tilde{b}^{ij})$.  (To be clear, at this point in the
analysis, the $b$'s and $\tilde{b}$'s do not separately describe
${\mathbb P}^9$'s, only the product of cones over copies of ${\mathbb P}^9.$
In the next step, when we take the intersection, the result will be 
equivalent to working in a single copy of ${\mathbb P}^9.$)
Furthermore, each of these two Pfaffians is equivalent
to $G(2,5)$ \cite{Donagi:2007hi}[section 4.2.2], and the remaining
superpotential term
\begin{displaymath}
 p_{i_1 i_2} \left( f^{i_1 i_2}(b) - \tilde{b}^{i_1 i_2}
\right)
\end{displaymath}
dictates that we should take the intersection of one copy of $G(2,5)$
with a linear rotation of the other copy of $G(2,5)$, following the
same analysis as described previously in this paper.

As a result, for $N=5$, for $r \gg 0$, we recover the same geometry --
a self-intersection of $G(2,5)$ with linear rotation -- that we recovered
in the previous GLSM, albeit in this GLSM, the geometry appears
wholly by nonperturbative effects as in \cite{Hori:2006dk},
rather than perturbatively.

Next we consider the phase $r \ll 0$.
In this phase, D-terms imply that not all of the $\varphi_i^a$.
$\tilde{\varphi}_i^{a'}$, and $p_{i_1 i_2}$'s can vanish.
The pertinent form for the superpotential is
\begin{displaymath}
W \: = \:
 \left( A(p)_{i_1 i_2} + \varphi_{i_1}^a \varphi_{i_2}^b J_{ab}
\right) b^{i_1 i_2} \: + \:
\left( C(p)_{i_1 i_2} + \tilde{\varphi}_{i_1}^{a'} \tilde{\varphi}_{i_2}^{b'}
J_{a' b'} \right) \tilde{b}^{i_1 i_2}.
\end{displaymath}
This is precisely an intersection of two PAXY models, each describing
a Pfaffian for the locus of rank $N-3$ of a matrix ($A(p)$, $C(p)$) in the
projective space defined by the $p$'s,
as described in \cite{Hori:2011pd} and \cite{Jockers:2012zr}[section 3.5].  
These were described
in those references as $Sp(N-3)$ gauge theories with a superpotential of the
form
\begin{displaymath}
b^{i j} \left( A_{i j} - x_i^a x_j^b J_{ab} \right),
\end{displaymath}
where the $b^{i j}$ are chiral superfields acting as Lagrange
multipliers, and the $x_i^a$ are introduced to enforce the rank constraint
on the antisymmetric matrix $A$.  The theory here in the $r \ll 0$ phase is
explicitly an intersection of two such PAXY models for Pfaffians,
as analyzed previously in section~\ref{sect:int-pfaff}.

In the case $N=5$, from \cite{Donagi:2007hi}[section 4.2.2], each
Pfaffian can be interpreted as a copy of $G(2,5)$, or more precisely a dual,
so we immediately recover the geometric interpretation of the
$r \ll 0$ phase of the previous GLSM for this geometry, as expected.
In the previous GLSM, the $r \ll 0$ phase was understood nonperturbatively
using the methods of \cite{Hori:2006dk}; here, by contrast,
the phase (in the dual GLSM) can be understood entirely perturbatively.

Using the parametrization of the maximal torus as outlined in section \ref{sec:int-grass-rot} one finds that the singular points in the moduli space match upon identification of the FI-$\theta$ parameters of the theory and its dual.

\subsection{Dualization on a single factor}

In the previous subsection, we analyzed the result of dualizing on
both $SU(2)$ factors.  For completeness, here we briefly outline the
analysis of a duality on a single $SU(2)$ factor.

Following the previous analysis, the dual GLSM has gauge group
\begin{displaymath}
\frac{
U(1) \times Sp(N-3) \times SU(2)
}{
{\mathbb Z}_2 \times {\mathbb Z}_2
},
\end{displaymath}
with 
\begin{itemize}
\item $N$ chiral multiplets $\varphi_i^a$ in the $({\bf N-3},{\bf 1})_{-1}$
representation,
\item $(1/2)N(N-1)$ chiral multiplets $b^{ij}=-b^{ji}$ in the $({\bf 1},{\bf 1})_{2}$ representation,
\item $N$ chiral multiplets $\tilde{\phi}^j_{a'}$ in the
$({\bf 1},{\bf 2})_1$ representation,
\item $(1/2)N(N-1)$ chiral multiplets $p_{ij}=-p_{ji}$ in the
$({\bf 1},{\bf 1})_{-2}$ representation,
\end{itemize}
with superpotential
\begin{eqnarray*}
W & = &
\sum_{i_1 < i_2}
\left( A(p)_{i_1 i_2} + \varphi_{i_1}^a \varphi_{i_2}^b J_{ab} \right)
b^{i_1 i_2} \: + \:
\sum_{i_1 < i_2}
C(p)_{i_1 i_2} \tilde{\phi}^{i_1}_{a'} \tilde{\phi}^{i_2}_{b'}
\epsilon^{a' b'}, \\
& = &
\sum_{i_1 < i_2} p_{i_1 i_2} \left( f^{i_1 i_2}(b) - \tilde{B}^{i_1 i_2}
\right) \: + \:
\varphi_{i_1}^a \varphi_{i_2}^b J_{ab} b^{i_1 i_2}.
\end{eqnarray*}
where $\tilde{B}$ is defined as previously.

Let us look in more detail at the gauge group and the quotient. First
write $\alpha\in U(1)$, $g_1\in USp(N-3)$, $g_2\in SU(2)$. Then:
\begin{displaymath}
  p_{ij}\rightarrow\alpha^{-2}p_{ij} \quad
  b^{ij}\rightarrow\alpha^2 b^{ij} \quad \varphi_i\rightarrow
  \alpha^{-1}g_1\varphi_i\quad \tilde{\phi}^j\rightarrow \alpha
  g_2\phi^j.
\end{displaymath}
Assigning the FI parameter $r$ to $U(1)$ the $U(1)$ D-term is
\begin{displaymath}
  2\sum_k (-|p_k|^2+|b_k|^2)+\sum_i (-|\varphi_i^a|^2+|\tilde{\phi}^j_b|^2)=2r
\end{displaymath}
This means that at $r\gg0$ not all $\tilde{\phi}$ and $b$ are allowed
to vanish whereas at $r\ll0$ not all $p$ and $\varphi$ are allowed to
vanish simultaneously. This is weaker than in the original model: here
the D-terms allow for branches where $b=0$ {\em or} $\tilde{\phi}=0$
at $r\gg0$ and $p=0$ {\em or} $\varphi=0$ at $r\ll0$. This looks like
there can be a mixture of strongly coupled and weakly coupled
components in either phase. Such phenomena have also been observed in
\cite{Hori:2016txh}.

Let us first consider the phase $r \gg 0$.  In this phase, the $U(1)$
D-terms imply that not all the $b^{ij}$ or $\tilde{\phi}^j_{a'}$ can vanish.
In this phase, it is most useful to write the superpotential in the form
\begin{displaymath}
W \: = \:
\sum_{i_1 < i_2} p_{i_1 i_2} \left( f^{i_1 i_2}(b) - \tilde{B}^{i_1 i_2}
\right) \: + \:
\varphi_{i_1}^a \varphi_{i_2}^b J_{ab} b^{i_1 i_2}.
\end{displaymath}

The first terms, $p(f(b) - \tilde{B})$,
appear to give a perturbative realization of
the intersection of the Pl\"ucker embedding of $G(2,N)$ (via the $\tilde{\phi}$)
with a rotation of the projective space ${\mathbb P}^{(1/2) N (N-1) -1}$
(defined by the $b$'s).

The second terms, $[\varphi,\varphi]_{ij} b^{ij}$, can be interpreted
nonperturbatively.  Working in a Born-Oppenheimer approximation on
${\mathbb P}^{(1/2)N (N-1)-1}$, this is a mass term for the $\varphi$'s.
Following the same logic as \cite{Hori:2006dk} and previously,
there are no vacua where
the matrix $[\varphi,\varphi]$ has rank $N-1$, only where it has rank
$N-3$ or less.  In this fashion, we find that the second terms dictate
an intersection with a Pfaffian variety in ${\mathbb P}^{(1/2)N(N-1)-1}$.

For $N=5$, this Pfaffian is the same as the Grassmannian $G(2,5)$
\cite{Donagi:2007hi}[section 4.2.2].  As a result, this phase
describes the self-intersection of $G(2,5)$ with a linear rotation,
matching the result for the duality frames, albeit here with a mix
of perturbative and nonperturbative analysis.

Next, let us consider the other phase, $r \ll 0$.  In this phase,
not all the $\varphi_i^a$ or $p_{ij}$ can vanish.  The superpotential
is most usefully written in the form
\begin{displaymath}
W \: = \: \sum_{i_1 < i_2}
\left( A(p)_{i_1 i_2} + \varphi_{i_1}^a \varphi_{i_2}^b J_{ab} \right)
b^{i_1 i_2} \: + \:
\sum_{i_1 < i_2}
C(p)_{i_1 i_2} \tilde{\phi}^{i_1}_{a'} \tilde{\phi}^{i_2}_{b'}
\epsilon^{a' b'}.
\end{displaymath}

In this phase, the first set of terms, $(A(p) + [\varphi, \varphi]) b$,
appear to give a perturbative realization of an intersection.
For $N=5$, this is the intersection of 
a Grassmannian $G(2,5)$ (whose baryons are $[\varphi \varphi]$) and
a rotation of the projective space ${\mathbb P}^{(1/2) N (N-1)-1}$
(defined by the $p$'s).

The second set of terms, $C(p) \tilde{B}$, can be interpreted
nonperturbatively.  Working in a Born-Oppenheimer approximation on
${\mathbb P}^{(1/2) N (N-1)-1}$, this is a mass term for the
$\tilde{\phi}$'s.  Following the same logic as
\cite{Hori:2006dk} and previously, there are no vacua where the
matrix $\tilde{B}^{ij}$ has rank $N-1$, only where it has
rank $N-3$ or less.  In this fashion, we find that the second terms
dictate an intersection with a Pfaffian variety in
${\mathbb P}^{(1/2) N (N-1)-1}$.

For $N=5$, for the same reasons as above, the Pfaffian coincides with
the (dual) Grassmannian, and so again we recover the same geometry
in the phase $r \ll 0$ as in other phases, albeit again via two
contributions, one perturbative, the other nonperturbative. The Coulomb branch analysis confirms that also the singular loci in the moduli space match.

\section{Other Kanazawa Calabi-Yaus}
\label{sect:kanazawa-exs}

The intersection of two copies of $G(2,5)$, discussed above,
is one of several examples of Calabi-Yau threefolds constructed
in \cite{kan}, listed there as $X_{25}$.  
Some of the examples discussed there are already
in the physics literature:  the example labelled
$X_{14}$ is R{\o}dland's model
\cite{rodland98,Hori:2006dk}, and $X_{5,7,10,13}$ were dscussed in 
\cite{Hori:2013gga}.  In this section, we will discuss physical
realizations of the remaining examples.

\subsection{$Y_{5}$, $Y_{10}$}

In \cite{kan} Kanazawa gives two examples (examples 2.17 and 2.18, labelled
there $Y_{10}$ and $Y_5$, respectively) 
of Calabi-Yaus that are complete intersections of Pfaffians 
with a hypersurface of degree $d$
in a weighted projective space of dimension $M-1$
and weights $q_k$. 
We propose that 
both examples can be realized in terms of a GLSM with gauge group 
$(U(1)\times USp(2))/\mathbb{Z}_2\simeq U(2)$. The matter content is
\begin{itemize}
\item $M$ chiral superfields $\phi_k$  in representation ${\bf 1}_{2q_k}$,
\item 5 chiral superfields $x_i^a$ in representation ${\bf 2}_1$,
\item 10 chiral superfields $b^{ij} = - b^{ji}$ in 
representation ${\bf 1}_{-2}$,
\item 1 chiral superfield $q$ in representation ${\bf 1}_{-2d}$.
\end{itemize}
The superpotential is
\begin{displaymath}
W \: = \: b^{ij} \left( A(\phi)_{ij} - x_i^a x_j^b \epsilon_{ab}
\right) \: + \: q f_d(\phi),
\end{displaymath}
where $A(\phi)$ is a skew-symmetric $5 \times 5$ matrix (whose rank
two locus defines the Pfaffian), with entries of charge $2$ under
the $U(1)$, and $f_d(\phi)$ is a polynomial of degree $d$
(hence charge $2d$). 
In the follwoing we will assume that the Calabi-Yau condition is satisfied.

Then there is a large volume phase $r\gg0$ where the D-terms imply that $p_{1,\ldots,M}$ and $x_{1,\ldots,5}$ are not allowed to vanish simultaneously and the matrix $x_i^a$ has rank $2$. The vacuum is a complete intersection of a Pfaffian with a degree $d$ hypersurface given by
\begin{displaymath}
  A_{ij}(\phi_k)- x_i^a x_j^b \epsilon_{ab}=0\qquad f_d(\phi_k)=0\qquad q,b^{ij}=0.
\end{displaymath}
The geometry is smooth if $\frac{\partial A_{ij}}{\partial \phi_k}=0$, $\frac{\partial f_d}{\partial \phi_k}=0$ do not have any solution other than $\phi_{1,\ldots,M}=0$. 

In the $r\ll 0$-phase $q$ and $b^{ij}$ are not allowed to vanish simultaneously. There is a vacuum where $\phi_k$, $x$ and $b^{ij}$ are zero and $p$ obtains a VEV. In this case the smoothness condition for $f_d$ is relevant. The $U(1)$-symmetry is broken to $\mathbb{Z}_d$.  Small fluctuations of the $\phi_k$ generate a potential and we find a Landau-Ginzburg orbifold. However, this is the whole story. There there is also non-perturbative realization of a complete intersection in a Grassmannian, similar to the Hori-dual where the $SU(2)$ is unbroken.

To further investigate this, let us consider the Hori dual of this model, which has the same non-abelian factor and the $b^{ij}$ are gone, while the $x_i$ have the opposite $U(1)$-charges compared to the original theory. The superpotential is
\begin{displaymath}
  W=A^{ij}(\phi_k)x_i^ax_j^b\epsilon_{ab}+qf_d(\phi_k).
\end{displaymath}
Now consider the $\tilde{r}\ll0$ phase where $q$ and $x$ are not allowed to vanish simultaneously. For $\phi_k=0$ there is a vacuum governed by
\begin{displaymath}
  A^{ij}_kx_i^ax_j^b\epsilon_{ab}=0,
  \end{displaymath}
This is a complete intersection of codimension $5$ in $G(2,5)$, which is an elliptic curve (see \cite{Hori:2013gga} for a discussion). This breaks the gauge symmetry completely. If $q$ gets a non-trivial VEV the $U(1)$ gets broken to $\mathbb{Z}_d$ and fluctuations around the vacuum state generate a Landau-Ginzburg potential $f_d$. However, this belongs to a different branch as the Grassmannian, since the $U(1)$ D-term reduces to $-2d|q^2|-||x||^2=\tilde{r}$, so there does not seem to be a unique vacuum. However, if we set $x=0$ there is such a vacuum and the phase looks like a Landau-Ginzburg model. So, this phase has two components, one where the $\phi_k$ are zero (but not necessarily $q$ which seems to have a non-compact direction) that looks like an elliptic curve, and one where $x=0$ and the phase looks like a Landau-Ginzburg model. Phases where the D-term and F-terms equations admit more than one branch of solutions have already been observed in \cite{Hori:2013gga,Hori:2016txh} and have been linked to ``bad'' hybrid models in the sense of \cite{Aspinwall:2009qy}.

In the next two subsections we will give further details for each of
these two models.

\subsubsection{$Y_{10}$}

This example is a Calabi-Yau threefold, constructed mathematically
as the intersection of a Pfaffian associated to
${\cal E} \equiv {\cal O}^5 \rightarrow {\mathbb P}^7_{[1^7,2]}$,
(in the sense that the matrix defining the Pfaffian is a section of
$\wedge^2 {\cal E}(t)$ for some $t$,)
with a quartic hypersurface in ${\mathbb P}^7_{[1^7,2]}$.
It is discussed in \cite{kan}[example 2.17].

We propose that the corresponding GLSM is given as follows.
Starting off with a gauge group 
$$U(2) = \frac{ U(1)\times SU(2) }{ {\mathbb Z}_2 },$$ 
the matter content is
\begin{itemize}
\item 7 chiral superfields $\phi_i$ in representation ${\bf 1}_2$,
\item 1 chiral superfield $\phi_8$ in representation ${\bf 1}_4$, 
forming the remaining homogeneous
coordinate on ${\mathbb P}^7_{[1^7,2]}$,
\item 5 chiral superfields $x_i^a$ in representation ${\bf 2}_1$,
\item 10 chiral superfields $b^{ij} = - b^{ji}$ in representation
${\bf 1}_{-2}$,
\item 1 chiral superfield $q$ in representation ${\bf 1}_{-8}$,
\end{itemize}
and superpotential
\begin{displaymath}
W \: = \:
b^{ij} \left( A(\phi)_{ij} - x_i^a x_j^b \epsilon_{ab} \right)
\: + \:
q f_4(\phi),
\end{displaymath}
where $A(\phi)$ is a skew-symmetric $5 \times 5$ matrix
(whose rank two locus defines the Pfaffian), with entries of charge $2$
under the $U(1)$, and $f_4(\phi)$ is a generic polynomial of degree four
(hence charge $8$).

The Calabi-Yau condition is satisfied: $-2\cdot 7-4\cdot 1+8\cdot 1-3\cdot 2\cdot 5+10\cdot 4=0$. Now we implement the $\mathbb{Z}_2$ quotient to rewrite the gauge group as $U(2)$. With $\alpha\in U(1)$, $g\in SU(2)$ and $\tilde{g}=\alpha g\in U(2)$ the fields $\phi_{1,\ldots,7}$ transform in the $\det$-representation and $\phi_8$ in the $\det^2$-representation. The $x$-fields transform in the fundamental and $q$ and $b^{ij}$ are in the $\det^{-4}$ and $\det^{-2}$ representations, respectively.

The entries of the matrix $A(p)$ are of degree $2$ in the $\phi_k$, 
{\it i.e.} linear in $\phi_{1,\ldots,7}$ while $\phi_8$ does not appear at all.
In the $r\gg0$ phase we find a geometry as described above. To see that the geometry is smooth we note that the only solution of $\frac{\partial f_4}{\partial \phi_8}$ is if all $\phi_k$ are zero which is excluded by the D-terms and F-terms. Note that the D-terms and F-terms, together with the genericity assumptions also exclude the problematic locus where $\phi_{1,\ldots,7}=0$ but $\phi_8\neq 0$ where the rank of $A(p)$ would drop to zero and there would be a singularity. The $r\ll0$ phase is as described above.

Reference \cite{kan} also says that this model coincides mathematically
with a double covering
of a Fano threefold.  We have not seen this description directly
in physics, but it would be interesting to do so.

Computing the effective potential on the Coulomb branch, one finds two conifold points
\begin{displaymath}
  e^{-t}=-\frac{1}{64}\frac{1}{(1-\omega)^5}, \; -\frac{1}{64}\frac{1}{(1+\omega^2)^5}\qquad \omega=e^{\frac{2\pi i}{10}}.
  \end{displaymath}
This, together with the topological data given in \cite{kan} can be found under the label AESZ 51 in the Calabi-Yau operator database \cite{vanstraten}.

\subsubsection{$Y_5$}

This example is a Calabi-Yau threefold, constructed mathematically
as the intersection of a Pfaffian associated to 
${\cal O}^5 \rightarrow {\mathbb P}^7_{[1^6,2,3]}$
with a degree 6 hypersurface in ${\mathbb P}^7_{[1^6,2,3]}$.
It is described in \cite{kan}[example 2.18].
(Note however that \cite{kan} did not verify the existence of smooth
examples of this form.)

We propose that this model is described by the following GLSM,
with gauge group 
$$U(2) = \frac{ U(1) \times SU(2) }{ {\mathbb Z}_2 },$$
and matter
\begin{itemize}
\item 6 chiral superfields $\phi_i$ in representation ${\bf 1}_2$,
\item 1 chiral superfield $\phi_7$ in representation ${\bf 1}_4$,
\item 1 chiral superfield $\phi_8$ in representation ${\bf 1}_6$,
\item 5 chiral superfields $x_i^a$ in representation ${\bf 2}_1$,
\item 10 chiral superfields $b^{ij} = - b^{ji}$ in 
representation ${\bf 1}_{-2}$,
\item 1 chiral superfield $q$ in representation ${\bf 1}_{-12}$.
\end{itemize}
and superpotential
\begin{displaymath}
W \: = \: b^{ij} \left( A(\phi)_{ij} - x_i^a x_j^b \epsilon_{ab}
\right) \: + \: q f_6(\phi),
\end{displaymath}
where $A(\phi)$ is a skew-symmetric $5 \times 5$ matrix (whose rank
two locus defines the Pfaffian), with entries of charge $2$ under
the $U(1)$, and $f_6(\phi)$ is a polynomial of degree six
(hence charge $12$).
The sum of the $U(1)$ charges vanishes, consistent with the Calabi-Yau
condition, and the superpotential above is explicitly gauge-invariant. Note that $\phi_{7,8}$ do not appear in $A(\phi)$. The analysis of the phases is indicated as above.

The Coulomb branch analysis yields two singular points at
\begin{displaymath}
   e^{-t}=-\frac{1}{432}\frac{1}{(1-\omega)^5}, \;-\frac{1}{64}\frac{1}{(1+\omega^2)^5}\qquad \omega=e^{\frac{2\pi i}{10}}.
  \end{displaymath}
This, together with the topological data given in \cite{kan} can be found under the label AESZ 63 in the Calabi-Yau operator database \cite{vanstraten}.

\subsection{$X_9$}

This is a Pfaffian associated to
${\cal E} \equiv {\cal O}(2) \oplus {\cal O}(1)^2 \oplus {\cal O}^2 
\rightarrow {\mathbb P}^6_{[1^6,2]}$. 
It's isomorphic to
${\mathbb P}^5[3,3]$.

We propose that the corresponding GLSM has gauge group 
$U(1) \times SU(2),$ 
with matter
\begin{itemize}
\item 6 chiral superfields $\phi_i$ in representation ${\bf 1}_1$,
\item 1 chiral superfield $\phi_7$ in representation ${\bf 1}_2$,
\item 5 chiral superfields $x_i^a$ in representations
${\bf 2}_2$, ${\bf 2}_1$, ${\bf 2}_1$, ${\bf 2}_0$, ${\bf 2}_0$,
\item 10 chiral superfields $b^{ij} = - b^{ji}$ that are $SU(2)$
singlets of charges
\begin{displaymath}
-3, -3, -2, -2, -2, -1, -1, -1, -1, 0,
\end{displaymath}
\end{itemize}
and superpotential
\begin{displaymath}
W \: = \: b^{ij} \left( A(\phi)_{ij} - x_i^a x_j^b \epsilon_{ab} 
\right),
\end{displaymath}
where $A(\phi)$ is a $5 \times 5$ skew-symmetric matrix which should
be interpreted as a section of $\wedge^2 {\cal E}$.
(An example of such a matrix $A$, in fact a one-parameter family of
such matrices, is given in \cite{kan}.)
The sum of the $U(1)$ charges vanishes, consist with the Calabi-Yau
condition, and the superpotential above is explicitly gauge-invariant.

\section{Conclusions}

In this paper we have described GLSMs for several further exotic 
constructions.  We began by describing GLSM realizations of Veronese,
Segre, and diagonal embeddings, which can be useful building-blocks for
various constructions of Calabi-Yau.  We then described GLSMs for
intersections of Grassmannians, after linear rotations and Veronese
and Segre embeddings.  We then discussed the (non-birational) phases 
of the intersections of two Grassmannians with a linear rotation,
giving a physical realization of mathematical structures recently
discussed in \cite{bcp,or}.  We then concluded with GLSMs for some other
exotic Calabi-Yau constructions in \cite{kan}.

We observed at one point that the GLSM for the Veronese map coincides with
the PAXY model of \cite{Jockers:2012zr} describing the image as a 
(hyper)determinantal variety.  Now, PAXY and PAX models exist in
greater generality, in which fields $x$, $y$ are introduced to realize
a determinantal variety.  To our mind this begs the question of whether
those fields can be understood in terms of defining maps in a fashion analogous
to what we have seen here for the Veronese embeddings.  We leave this
matter for future work.

Analogous mathematical 
phenomena are described in \cite{manivel17} for spinor varieties,
and in \cite{pertusi} for cubic fourfolds.
Further, more exotic mathematical examples can also be
found in \cite{eckvan}[section 6], \cite{CO:To,tonoli2}.
It would be very interesting to construct corresponding GLSMs.

A further interesting class of examples to look at are Calabi-Yaus that do not have a point of maximal unipotent monodromy as discussed for instance in \cite{cv1,cv2,cv3}. This would translate into GLSMs that do not have geometric phases. In such cases localization techniques from the GLSM that do not rely on mirror constructions may be particularly useful for analyzing these Calabi-Yaus. 

Most known examples of non-abelian GLSMs have matter that transforms either in the fundamental or some one-dimensional representation. In section \ref{sec:nonab-ver} we have discussed GLSMs with matter in the ${\bf {\rm Sym}^d k}$ representation. It would be interesting to analyze GLSMs with exotic matter in further detail, as they will lead to interesting new examples of Calabi-Yau spaces.

Finally, the construction of intersecting Grassmannians seems to fit into a more general framework \cite{galkin-talk}. 
It would be interesting to see if this approach leads to a systematic construction of a larger class of new Calabi-Yaus via non-abelian GLSMs.

\section{Acknowledgements}

We would like to thank S.~Galkin, S.~Katz,
I.~Melnikov, D.~Morrison, 
T.~Pantev, J.~Rennemo, M.~Romo, S.~Sam, E.~Scheidegger and E.~Segal 
for useful conversations.
E.S. was
partially supported by NSF grants PHY-1417410 and PHY-1720321.

\end{document}